\documentclass[aps,preprint,nofootinbib,superscriptaddress]{revtex4}
\pdfoutput=1
\usepackage{color}
\definecolor{note_fontcolor}{rgb}{0.80078125, 0.80078125, 0.80078125}
\usepackage{verbatim}
\usepackage{slashed}
\usepackage{amssymb,amsmath,cancel}
\usepackage[colorlinks=true,linkcolor=blue,citecolor=blue,filecolor=blue,
urlcolor=blue]{hyperref}

\usepackage{graphicx}

\makeatletter

\newenvironment{lyxgreyedout}
  {\textcolor{note_fontcolor}\bgroup\ignorespaces}
  {\ignorespacesafterend\egroup}

\@ifundefined{textcolor}{}
{%
 \definecolor{BLACK}{gray}{0}
 \definecolor{WHITE}{gray}{1}
 \definecolor{RED}{rgb}{1,0,0}
 \definecolor{GREEN}{rgb}{0,1,0}
 \definecolor{BLUE}{rgb}{0,0,1}
 \definecolor{CYAN}{cmyk}{1,0,0,0}
 \definecolor{MAGENTA}{cmyk}{0,1,0,0}
 \definecolor{YELLOW}{cmyk}{0,0,1,0}
 }

\makeatother

\usepackage{bbm}                                                  
\newcommand{\nc}{\newcommand}
\nc{\beq}{\begin{equation}}  \nc{\eeq}{\end{equation}}
\nc{\bea}{\begin{eqnarray}}  \nc{\eea}{\end{eqnarray}}
\nc{\baa}{\begin{array}}     \nc{\eaa}{\end{array}}
\nc{\bit}{\begin{itemize}}   \nc{\eit}{\end{itemize}}
\nc{\ben}{\begin{enumerate}} \nc{\een}{\end{enumerate}}
\nc{\bce}{\begin{center}}    \nc{\ece}{\end{center}}
\nc{\bpm}{\begin{pmatrix}}   \nc{\epm}{\end{pmatrix}}
\nc{\bvt}{\begin{verbatim}}  \nc{\evt}{\end{verbatim}}
\def\half{\frac12}	%\def\half{{1\over2}}

\def\to{\rightarrow}

\def\gesim{\gtrsim}
\def\lesim{\lesssim}
\def\boldoverdot{\,{\raise6pt\hbox{\bf.}\!\!\!\!\>}}

\def\then{{\quad\Rightarrow\quad}}

\def\lcal{{\cal L}}
\def\mcal{{\cal M}}

\def\ocal{{\cal O}}

\def\wcal{{\cal W}}

\def\bBB{{\mathbbm B}}%\def\bBB{ \hbox{{\mysmallii I}}\!\hbox{{\mysmallii B}} }

\usepackage{bm}
\def\sigbf{{\bm\sigma}}		%	\def\sigbf{{\pmb{$\sigma$}}}
\def\ssb{spontaneous symmetry breaking}
\def\vev{vacuum expectation value}

\def\Tr{ \hbox{Tr}}

\def\diag{\hbox{\diag}}

\def\ev{\hbox{eV}}

\def\gev{\hbox{GeV}}
\def\tev{\hbox{TeV}}

\def\vevof#1{\left\langle #1 \right\rangle}

\def\doubleundertext#1{
{\undertext{\vphantom{y}#1}}\par\nobreak\vskip-\the\baselineskip\vskip4pt%
\undertext{\hbox to 2in{}}}
\def\inbox#1{\vbox{\hrule\hbox{\vrule\kern5pt
     \vbox{\kern5pt#1\kern5pt}\kern5pt\vrule}\hrule}}
\def\sqr#1#2{{\vcenter{\hrule height.#2pt
      \hbox{\vrule width.#2pt height#1pt \kern#1pt
         \vrule width.#2pt}
      \hrule height.#2pt}}}

\def\today{\ifcase\month\or
  January\or February\or March\or April\or May\or June\or
  July\or August\or September\or October\or November\or December\fi
  \space\number\day, \number\year}
\def\pmb#1{\setbox0=\hbox{#1}%
  \kern-.025em\copy0\kern-\wd0
  \kern.05em\copy0\kern-\wd0
  \kern-.025em\raise.0433em\box0 }

\def\pmbb#1{\setbox0=\hbox{#1}%
  \kern-.02em\copy0\kern-\wd0
  \kern.04em\copy0\kern-\wd0
  \kern-.02em\raise.03464em\box0 }
\def\up#1{^{\left( #1 \right) }}
\def\inv#1{\frac1{#1}}
\def\su#1{{SU(#1)}}
\def\ui{U(1)}
\def\sumprime_#1{\setbox0=\hbox{$\scriptstyle{#1}$}
  \setbox2=\hbox{$\displaystyle{\sum}$}
  \setbox4=\hbox{${}'\mathsurround=0pt$}
  \dimen0=.5\wd0 \advance\dimen0 by-.5\wd2
  \ifdim\dimen0>0pt
  \ifdim\dimen0>\wd4 \kern\wd4 \else\kern\dimen0\fi\fi
\mathop{{\sum}'}_{\kern-\wd4 #1}}
\def\sc#1#2{\Phi_{#1}\up{#2}}
\def\fer#1#2{\Psi_{#1}\up{#2}}
\def\bos#1#2{B_{#1}\up{#2}}
\def\vec#1#2{X_{#1}\up{#2}}

\def\mw{m_{\rm w}}

\def\lt{\tilde\ell}
\def\lb{\bar\ell}
\def\phid{\phi^\dagger}
\def\phit{\tilde\phi}

\def\xor{~;~~}

\def\oi{{7{\rm\hbox{-}I}}}
\def\oii{{7{\rm\hbox{-}II}}}
\def\oiii{{7{\rm\hbox{-}III}}}
\def\oiv{7}

\def\cw{c_{\rm w}}
\def\sw{s_{\rm w}}

\def\znbb{$0\nu\beta\beta$ }

\def\sc#1#2{\Phi_{#1}\up{#2}}
\def\fer#1#2{\Psi_{#1}\up{#2}}
\def\bos#1#2{B_{#1}\up{#2}}
\def\vec#1#2{X_{#1}\up{#2}}

\def\mw{m_{\rm w}}

\def\lt{\tilde\ell}
\def\lb{\bar\ell}
\def\phid{\phi^\dagger}
\def\phit{\tilde\phi}

\def\xor{~;~~}

\def\oi{{7{\rm\hbox{-}I}}}
\def\oii{{7{\rm\hbox{-}II}}}
\def\oiii{{7{\rm\hbox{-}III}}}

\def\cw{c_{\rm w}}
\def\sw{s_{\rm w}}

\def\aut{\,{\rm or}\,}
\def\td{\hbox{{\scriptsize $\left[\aut \sc01 \right]$}}}
\def\tdf{\hbox{{\scriptsize$\left[\aut \fer00 \right]$}}}
\def\tdv{\hbox{{\scriptsize$\left[\aut \vec01 \right]$}}}

\begin{document}

\global\long\def\abs#1{\left| #1 \right| }
\global\long\def\half{\frac{1}{2}}
\global\long\def\partder#1#2{\frac{\partial#1}{\partial#2}}
 \global\long\def\comm#1#2{\left[ #1 ,#2 \right] }

\global\long\def\Tr#1{\textrm{Tr}\left\{  #1 \right\}  }

\global\long\def\Imag#1{\mathrm{Im}\left\{  #1 \right\}  }

\global\long\def\Real#1{\mathrm{Re}\left\{  #1 \right\}  }

\global\long\def\db{\!\not\!\! D\,}

\global\long\def\gesim{\,{\raise-3pt\hbox{$\sim$}}\!\!\!\!\!{\raise2pt\hbox{$>$}
}\,}
\global\long\def\then{{\quad\Rightarrow\quad}}
\global\long\def\lcal{{\cal L}}
\global\long\def\mcal{{\cal M}}
\global\long\def\bBB{{\mathbbm B}}
\global\long\def\sigbf{\bm{\sigma}}
\global\long\def\gev{\hbox{GeV}}
\global\long\def\tev{\hbox{TeV}}
\global\long\def\vevof#1{\left\langle #1\right\rangle }
\global\long\def\up#1{^{\left( #1 \right) }}
\global\long\def\inv#1{\frac{1}{#1}}
\global\long\def\su#1{{SU(#1)}}
\global\long\def\ui{U(1)}

\begin{comment}
CAFPE-166/11,UG-FT-296/11,FTUV-11-0414,IFIC/11-66,UCRHEP-T519
\end{comment}

\title{Effective Lagrangian approach to neutrinoless double beta decay 
and neutrino masses}

\author{Francisco del Aguila}

\affiliation{CAFPE and Departamento de Fisica Teorica y del Cosmos, Universidad
de Granada, E\textendash{}18071 Granada, Spain}

\author{Alberto Aparici}

\affiliation{Departament de Fisica Teorica, Universitat de Valencia and IFIC,
Universitat de Valencia-CSIC, Dr.~Moliner 50, E-46100 Burjassot (Valencia),
Spain}

\author{Subhaditya Bhattacharya}

\affiliation{Department of Physics and Astronomy, University of California,
Riverside
CA 92521-0413, USA}

\author{Arcadi Santamaria$^2$}

\author{Jose Wudka$^3$}

\begin{abstract}
Neutrinoless double beta ($0\nu\beta\beta$) decay can in general 
produce electrons of either chirality, in contrast with the minimal 
Standard Model (SM) extension with only the addition of the Weinberg operator, 
which predicts two left-handed electrons in the final state. 
We classify the lepton number violating (LNV) effective operators with two
leptons of either chirality 
but no quarks, ordered according to the magnitude of their contribution to
\znbb decay. 
We point out that, for each of the three chirality assignments, $e_Le_L, e_Le_R$
and $e_Re_R$, 
there is only one LNV operator of the corresponding type to lowest order, and
these have
dimensions 5, 7 and 9, respectively.
Neutrino masses are always induced by these extra operators but can be delayed 
to one or two loops, depending on the number of RH leptons entering in the
operator. 
Then, the comparison of the \znbb decay rate and neutrino masses 
should indicate the effective scenario at work, which confronted with the LHC
searches 
should also eventually decide on the specific model  elected by nature. 
We also list the SM additions generating these operators upon integration 
of the heavy modes, and discuss 
simple realistic examples of renormalizable theories for each case.

%\begin{comment}
%\keywords{Neutrinos, neutrinoless double beta decay, effective Lagrangian, 
%doubly-charged scalars, LHC}
%\pacs{12.15.Ji, 12.60.Fr, 14.60.Pq, 14.80.Fd}
%\maketitle
%\end{comment}

\end{abstract}
\maketitle

\section{Introduction\label{sec:Introduction}}

The remarkable observation of neutrino oscillations (see
\cite{Nakamura:2010zzi} 
and \cite{Mohapatra:2005wg,GonzalezGarcia:2007ib} for recent reviews)
provided the
first direct evidence of physics beyond the Standard Model (SM),
these effects are best explained by providing the neutrinos
by small masses and appropriate mixing angles
\cite{Pontecorvo:1967fh,Maki:1962mu}
(see also \cite{Schwetz:2011zk} for a recent fit). In contrast
with the quark sector, however, neutrino masses are not
necessarily of the Dirac type, yet oscillation experiments are not
sensitive to determine whether neutrino
masses are of the Majorana type~\cite{Mohapatra:1998rq}. Fortunately there is
another process, neutrino-less double-beta ($0\nu\beta\beta$) decay (\cite{Racah:1937qq,Furry:1939qr}
and \cite{Vergados:2002pv} for a review)
that probes this property of the neutrino sector, and
has achieved sufficient sensitivity (see \cite{Avignone:2007fu,Barabash:2011fg,Elliott:2012sp}
for recent reviews) 
to provide interesting
constraints on the lepton-number violating (LNV) 
processes that can produce it. In this paper
we will be concerned with general properties of both
neutrino masses and \znbb decay; we will strive to provide 
model-independent description of the effects
that concern us, yet we will make connection with specific models
that provide concrete and important illustrations of the arguments
presented.

Since both neutrino masses and \znbb decay are low energy processes,
an effective Lagrangian approach
\cite{Coleman:1969sm,Weinberg:1978kz,Weinberg:1980wa,Polchinski:1983gv,Georgi:1994qn,Wudka:1994ny} 
is the proper starting
point of any model-independent discussion; in it, all virtual new
physics (NP) effects are parameterized by the coefficients, $C\up{n}$,
of the corresponding effective operators $\ocal\up{n}$. Explicitly,
\beq
\lcal = \lcal_{SM} + \sum_{n=5}^\infty \sum_i  
\left(\frac{C^{(n)}_i}{\Lambda^{n-4}} \ocal\up n_i + \mathrm{h.c.}\right)\, ,
\label{eq:effectiveL}
\eeq
where $n$ denotes the canonical dimension of the operator, 
$i$ labels the independent operators of a given order 
and $ \Lambda $ the NP scale. The $ \ocal_i\up n$ respect all the
local symmetries of the SM, but not necessarily
the global ones; when the NP contains several scales
$ \Lambda > \Lambda ' > \cdots $ the coefficients $C$ may contain
powers of $ \Lambda'/\Lambda $ (we will assume that the new physics
is weakly coupled and decoupling).

Several earlier papers~\cite{Babu:2001ex,Choi:2002bb,Engel:2003yr,deGouvea:2007xp} have followed this approach,
considering, however, only effective interactions that do not involve the SM
gauge bosons. Here we consider a different class of theories where the NP does
not couple directly to the quark sector~\footnote{Operators involving leptons and quarks with no gauge bosons
generate neutrino masses at 1 to 4 loops~\cite{deGouvea:2007xp,Babu:2010vp,Duerr:2011zd,Babu:2011vb}
and may receive enhancements from top Yukawa couplings.}, so that the effective interactions
involve only leptons and gauge bosons (coupling to the gauge bosons is generated
whenever the NP is not a SM gauge singlet). We will show that in this case one can provide a simple 
classification of the effects that concern us
in terms of only three operators, each of which
can be generated at tree level by different types of NP.
In the unitary gauge these operators give the vertices
$\nu_L\nu_L,~W e_R \nu_L$ and $ WW e_R e_R$
and have dimension 5,7 and 9, respectively. 

This allows for three scenarios wherein one of the
operators is generated at tree level and the others
via loops. A simple example
exhaustively considered in the  literature
has tree-level generated neutrino masses
via a high-scale see-saw
mechanism~\cite{Minkowski:1977sc,GellMann:1980vs,Yanagida:1979as,
Mohapatra:1979ia}, 
with effective $ W e_R \nu_L $ and $ WWe_R e_R $ vertices generated
radiatively. Here we concentrate on the complementary
situations where the other operators are 
generated at tree level and neutrino
masses are generated radiatively.
Though we will discuss the three cases
separately it is of course possible
for the NP to simultaneously
contain all of them. Indeed, in specific models, phenomenological
constraints  might
necessitate such complications; this is the case in 
realistic left-right (L-R) models \cite{Tello:2010am}.
However, the interesting
result of the analysis we present is that, whatever the complications,
they reduce to combinations of the three cases
discussed below whenever our assumptions are applicable.

The classification of NP contributions to 
\znbb decay is straightforward. The final state involves
two electrons of either chirality which may proceed directly 
from the effective vertex generated by the NP, or
from the $ W \nu_L e_L$ SM vertex.
As a result LNV effects
contributing to \znbb decay can be classified according
to the chirality of the final leptons. 
It is remarkable that, as we shall show,
{\it there is only one lowest-order operator for each 
of the three possible  chirality assignments exhibiting a notable
connection between chirality and dimension:}~\footnote{Notice that the $\ocal\up5$ lepton number (LN) assignment is 
opposite to that of 
$\ocal\up7$ and $\ocal\up9$ in order to preserve the usual convention for $\ocal\up5$.}
\begin{align}
&\ocal\up5 = (\overline{\tilde{\ell}_{L}}\phi)(\tilde \phi^\dagger \ell_{L}) , 
& {\rm for\ }&{\rm two\ LH\ leptons\ (LL)} ,  
\label{eq:lowestoperatorLL} \\ 
&\ocal\up7 = (\phid D^\mu \phit) (\phid \overline{e_R} \gamma_\mu \lt_L) ,
& {\rm for\ }&{\rm one\ LH\ lepton\ and\ one\ RH\ charged\ lepton\ (LR)},  
\label{eq:lowestoperatorLR} \\ 
&\ocal\up9 = \overline{e_R} e_R^c (\phid D^\mu \phit) (\phid D_\mu \phit) , 
& {\rm for\ }&{\rm two\ RH\ charged\ leptons\ (RR)} . 
\label{eq:lowestoperatorRR}
\end{align}
Here we omitted flavor indices and
denote the light scalar isodoublets by $\phi$, the
left-handed (LH) lepton isodoublets by $ \ell_L$, and
right-handed (RH) lepton isosinglets by $e_R$; we also use
$\tilde\phi=i\sigma_2\phi^*$,  and
$\tilde{\ell}_L=i\sigma_2\ell^c_L$.
The electric charge is the sum of the third component of isospin 
and the hypercharge, $Q = I_3 + Y$;
thus $\ell_L$ and $\phit$ have hypercharge $-1/2$, while
$\phi$ and $\tilde{\ell_L}$ have hypercharge $+1/2$. 

\begin{figure}
\includegraphics[width=0.95\columnwidth]{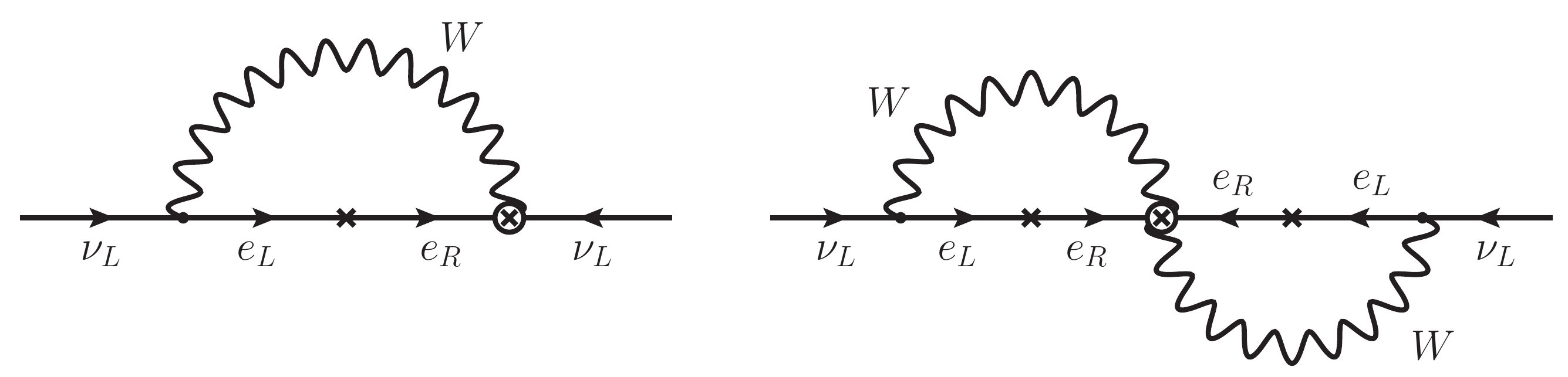}
\caption{One (left) and two (right) loop neutrino masses for LR and RR
operators, 
respectively. Arrows indicate fermion number flow. 
\label{fig:1L2Lmasses}}
\end{figure}

The operators in Eqs.~(\ref{eq:lowestoperatorLL}--\ref{eq:lowestoperatorRR}) 
also provide contributions to the neutrino
masses; the first one at
tree level, as first noted by Weinberg \cite{Weinberg:1979sa} (see also
\cite{Weldon:1980gi}), 
the other two, radiatively
at one and two loops, respectively; see Fig.~\ref{fig:1L2Lmasses}.
We will show that~\footnote{The dominant contributions 
obtained in specific models are proportional to $\Lambda^{-1} $ since the lowest
dimension operator describing neutrino masses is $\ocal\up5$, and this operator
has dimension 5. Once LN is broken, $\ocal\up5$ will always be generated, at one or
two 
loops (barring model dependent cancellations), even if it does not appear at
tree level; 
this happens in the models we will consider that generate $\ocal\up7$ or $\ocal\up9$
at tree level. 
Deriving these estimates requires care;
in particular, estimates obtained using the unitary gauge are
often not reliable.} 
(here $a,b$ are family indices)
\begin{align}
&(m_\nu)_{ab} \propto \frac{v^2}{\Lambda} C^{(5)*}_{ab}, 
& {\rm for\ LL} ,  
\label{al:massesLLcorrect} \\ 
&(m_\nu)_{ab} \propto \frac{v}{16 \pi^2 \Lambda} \left(m_a C\up7_{ab} + 
m_b C\up7_{ba}\right) ,
& {\rm for\ LR} , 
\label{al:massesLRcorrect} \\
&(m_\nu)_{ab} \propto \frac{1}{(16 \pi^2)^2 \Lambda} m_a C\up9_{ab} m_b , 
& {\rm for\ RR} ,
\label{al:massesRRcorrect}
\end{align}
where the vacuum expectation value (VEV) $\langle \phi\rangle \equiv v \simeq 
174~\mathrm{GeV}$
is the electroweak symmetry breaking parameter and $m_{a,b}$ the corresponding charged lepton masses.
It is worth noting that the contributions from $ \ocal\up{7,9}$
to $ m_\nu $ have a natural hierarchy derived from the 
charged lepton mass factors and may be used to generate
textures naturally. We will discuss this issue in a forthcoming
publication.

In all cases the neutrino masses
vanish when the scale of NP
$ \Lambda \to \infty$, as required 
by the decoupling theorem~\cite{Appelquist:1974tg}. 
This implies that the neutrino oscillation data imposes a {\it upper}
bound on $ \Lambda $, while the limit from \znbb decay provides a lower bound on this scale. It
also follows that the NP cannot be flavor blind, so that
specific models in general will be constrained, not only
by the neutrino and \znbb data, but also by lepton-flavor
violation (LFV) constraints~\cite{delAguila:2011gr}.

If the NP generates 
any one of these operators at tree level,
the remaining two will be generated at one or two loops; 
when $ \ocal\up5$ is generated radiatively the neutrino
masses are proportional to the coefficients of the 
tree-level generated operator and are, in this sense, 
predictable. 
In a companion paper~\cite{delAguila:2011gr} we provide a realistic and highly
constrained
model that illustrates this scenario, having
only $ \ocal\up9$  generated at tree level
while $ \ocal\up7,~\ocal\up5 $ appear,
respectively at one and two loops. In particular,
neutrino masses are generated at two loops
(see Eq. (\ref{al:massesRRcorrect}));
the wide literature on radiative neutrino masses
provides many additional examples (see for instance
\cite{Zee:1980ai,Zee:1985id,Babu:1988ki,Ma:2006km}).

We will see that current data on \znbb decay implies that the scales associated with
NP generating $ \ocal\up5,~\ocal\up7$ and $ \ocal\up9$
at tree level are, respectively
$ \Lambda\ > 10^{11}\left|C_{ee}\up5\right|\ \tev, 
10^2\left|C_{ee}\up7\right|^{1/3}\ \tev$, and 
$\left|C_{ee}\up9\right|^{1/5}\ \tev$. 
For first type of NP 
we expect all collider effects to be negligible;
while we expect that the NP  responsible for $ \ocal\up9 $
will be probed at the LHC \cite{delAguila:2011gr}. The intermediate case of
$ \ocal\up7$ may or may not have collider signatures,
depending on the details of the model; we examine one
such case below.

In next section we shall define our notation and using the
effective Lagrangian approach
we classify the lowest order interactions (operators) invariant
under the SM gauge group
with two external leptons but no quarks, 
mediating \znbb decay
(it turns out that such operators have equal number of covariant 
derivatives and RH leptons). 
The mechanism for generating neutrino masses is discussed  in
Section \ref{sec:neutrinomasses} for the three different cases 
and compared to the corresponding \znbb decay
amplitude. 
In Section  \ref{sec:mediators} 
we identify the new particle additions generating
those operators at tree level
after integrating them out. 
A more detailed exposition is presented in the appendix. 
Whereas explicit (renormalizable) models are given for each
scenario in Section \ref{sec:models}.
The phenomenological implications for LFV processes 
and LHC searches are reviewed in Section
\ref{sec:phenomenology}. The last
section is devoted to conclusions. 
As indicated, technical details are
collected in the appendix.

\section{Lowest order effective operators contributing to \znbb decay}

The effects of NP below its characteristic scale $\Lambda$ are 
adequately described by an effective theory involving the light 
fields and preserving the unbroken local symmetries;
the smaller (compared to $ \Lambda $)
the characteristic energy scale of the relevant
processes, the higher the accuracy of such a description.
This justifies parameterizing \znbb decay, whose 
effective scale is some tens of MeV, by a set of effective operators 
describing NP above the electroweak scale. These operators  must violate 
LN, 
as does \znbb decay and the observed neutrino masses, 
which we assume to be of Majorana type (we also assume there are 
no light RH neutrinos). 

In this section we will classify the lowest 
order local operators mediating 
\znbb decay. As mentioned in the introduction,
this has been addressed previously in the 
literature, often including quark fields~\cite{Babu:2001ex,Choi:2002bb,Engel:2003yr,deGouvea:2007xp}
(see also \cite{Babu:2010vp,Babu:2011vb} for recent models realizing operators with quark fields).
In contrast to these papers, we shall assume that the NP does
not
couple directly to the quark sector, so that all
quark interactions are mediated by the  
electroweak gauge bosons. 
In the appendix we provide the methodology for constructing the 
effective operators to any given order, discussing in detail the dominant ones 
for each final electron chirality.  

In order to write down the basis of effective operators we first have 
to fix 
the light field content and the symmetries the operators must satisfy. 
We shall restrict ourselves to the SM fields and local symmetries, 
although allowing for more that one light scalar doublet, 
as this gives a few more alternatives and may 
simplify explicit realistic realizations, as we shall illustrate later on. 
Operators contributing to \znbb decay 
must involve two leptons of either chirality, $\ell_L$ or $e_R$, 
and a number of scalar doublets, $\phi$, to make the product invariant under 
$SU(2)_L\times U(1)_Y$ transformations. Besides, they can have covariant 
derivative ($D_\mu$) insertions, which do not change the field quantum numbers. 
We will be mainly interested in the lowest order operators of each class, 
and in the heavy particles whose virtual effects can generate them
(assuming only renormalizable couplings). 

Before proceeding we note, that in some cases the number 
of different light scalar doublets $\phi_i$ in the theory matters,
since for $ i>1 $ qualitatively new operators become possible; 
this is related to the appearance of
additional physical scalars. When more than one light doublet
is present one must consider
all possible scalar flavor assignments to all allowed
operator structures. 
For example, in the RR case we have an operator
of dimension 7, analogous to $\ocal\up9$, but 
with no covariant derivatives:
$\overline{e_R} e_R^c (\phid_i \phit_j)^2$.
This field product vanishes if $i=j$ because 
the scalar product $\phid_i \phit_j$ of two isodoublets is antisymmetric;
however, if we have two light scalar doublets, the operator
does not vanish. Such 
a theory will contain light charged scalars $H^\pm$; upon
\ssb, $\phid_i \phit_j \propto H^- + \cdots $, so that
operator generates the coupling
$\overline{e_R} e_R^c H^- H^-$ and
will contribute to \znbb decay if the physical charged Higgs $H^-$ 
couples to quarks.
Similarly, the number of sets of possible heavy excitations that
generate the effective operators after being integrated out
increases with the number of independent light scalar doublets. 

We will proceed as follows. In the classification of 
independent operators we will for the most part
assume there is only one SM Higgs 
doublet. We will do the same for the listing of the SM additions 
(that is, the heavy excitations)
generating these operators; in the appendix we also 
discuss the differences when there is more than one light 
scalar isodoublet.
In the model generating $\ocal\up7$ 
below (Section \ref{sec:LR-model}) 
we make use of two light scalar doublets for illustration, 
and simplicity; in that particular case
the presence of a discrete symmetry guarantees that 
there is only one lowest-order effective operator 
involving two leptons, one LH and one RH,
despite having two doublets (technically this
is because a covariant derivative  and a gamma
matrix are both required to match chirality and
Lorentz  indices).

\subsubsection{LNV operators with two LH leptons}

\begin{figure}
\begin{centering}
\includegraphics[width=0.55\columnwidth]{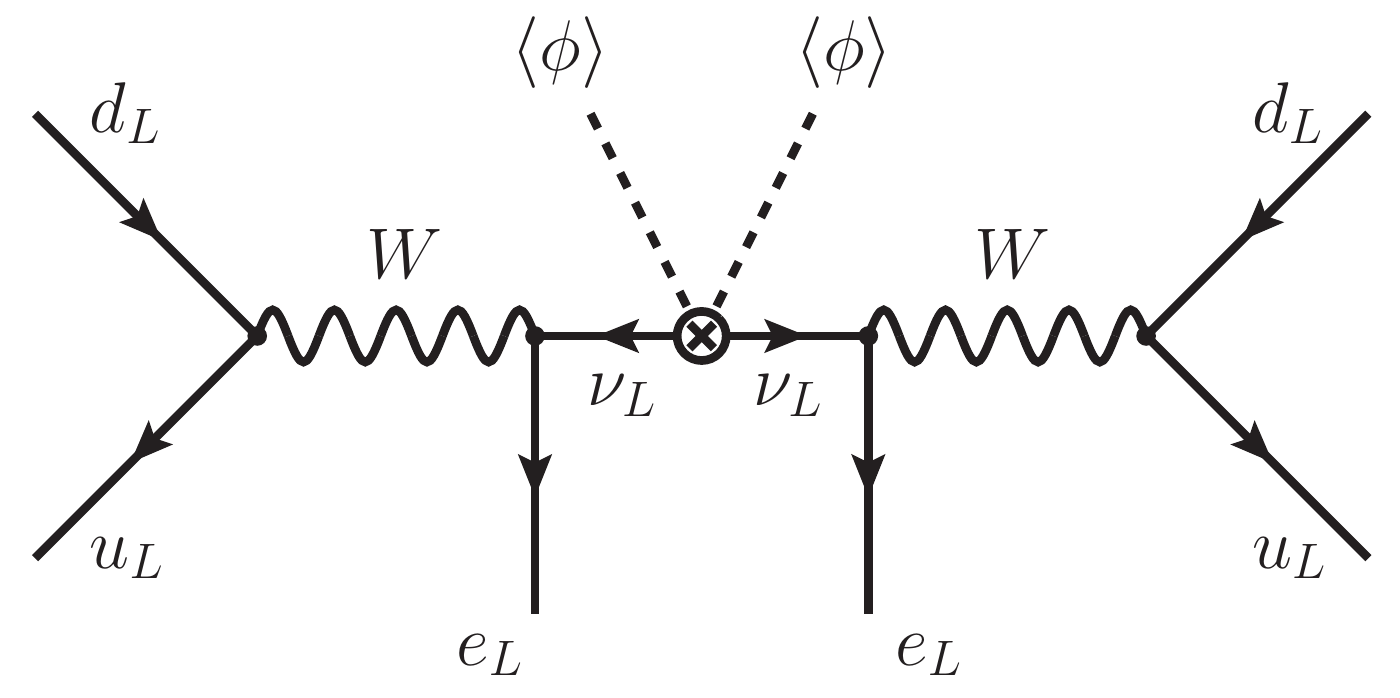}
\par\end{centering}
\caption{Tree-level diagram mediated by $\ocal\up5$ (light neutrino masses) 
contributing to \znbb decay.
\label{fig:0nu2betaSM}}
\end{figure}

The lowest-order operator of this type is 
the only invariant dimension 5 (Weinberg) operator 
\cite{Weinberg:1979sa,Weldon:1980gi} displayed in
Eq.~(\ref{eq:lowestoperatorLL}).
The corresponding term in the Lagrangian is
\beq
\frac{C\up5_{ee}}{\Lambda} \ocal_{ee}\up5 = \frac{C\up5_{ee}}\Lambda
\overline{\tilde\ell_{e\,L}}\phi \tilde\phi^\dagger \ell_{e\,L} \rightarrow 
-\frac{ v^2 C\up5_{ee}}\Lambda \overline{\nu^c_{e\,L}}\nu_{e\,L}+\cdots
= -\frac{1}{2}(m_\nu)^*_{ee}  \overline{\nu^c_{e\,L}}\nu_{e\,L}+\cdots
\, , 
\label{eq:operatorll5}
\eeq
where the isodoublet corresponds to the first generation
(as will be the case throughout this section).
In general there are corresponding contributions for all
lepton flavors, see Section \ref{sec:neutrinomasses},
which provide  Majorana masses to all the SM neutrinos after 
electroweak symmetry breaking. 
This operator contributes to \znbb decay by transforming 
the two neutrinos into two LH electrons by the emission of 
two $W$ gauge bosons (see Fig. \ref{fig:0nu2betaSM}). 

All LNV operators of dimension 6
involve quarks and violate baryon number by $ \pm1 $ unit;
they also conserve $B-L$ and will not contribute to \znbb decay.
There are, however, 
three independent operators of dimension 7 that do contribute to 
\znbb decay:
\begin{eqnarray}
\ocal\up{\oi} &=& (\overline{D_\mu \ell_L}\tilde{\phi})(\phid D^\mu \lt_L)\,,\cr
\ocal\up{\oii} &=& (\overline{\ell_L} D_\mu \lt_L)(\phid D^\mu \phit) \,,\cr
\ocal\up{\oiii} &=& (\overline{\ell_L}\tilde{\phi}) \partial_\mu (\phid D^\mu
\lt_L) \,
\label{eq:operatorsll7}
\end{eqnarray}
(see the appendix).
One may think that the operators in Eq. (\ref{eq:operatorsll7}) 
generate important contributions to LNV processes; 
this, however, is not the case.
The reason is that {\em any} model generating (\ref{eq:operatorsll7})
at tree level {\em necessarily} also generates $ \ocal\up5 $ in Eq.
(\ref{eq:operatorll5}) at tree level; all these operators
contribute to all the processes we are interested in, but those from 
(\ref{eq:operatorsll7}) will always be suppressed by two additional powers
of $ \Lambda $ and are
subdominant. 
Accordingly, we will ignore these operators 
in the phenomenological analysis.

The limit on \znbb decay, for instance, from  $^{76}$Ge 
\cite{KlapdorKleingrothaus:2000sn,Aalseth:2002rf} (see also
\cite{Barabash:2011fg,Avignone:2007fu} for recent reviews) is usually expressed in terms
of the electron-electron element of the neutrino mass matrix
\begin{equation}
|(m_\nu)_{ee}| < 0.24  - 0.5\ \ev\, .
\label{eq:0nu2betaLLnumasslimit}
\end{equation}
This constraint on $(m_\nu)_{ee}$
is consistent with the neutrino mass limits from oscillations, cosmology and tritium beta decay.
When \eqref{eq:operatorll5} is used to express $(m_\nu)_{ee}$ in terms of $C_{ee}\up5$, the following
restriction 
on $\Lambda$ is obtained
\begin{equation}
 \frac{\Lambda}{|C_{ee}\up5|} > 10^{11}\ \tev\, .
\label{eq:LambdaC5ee}
\end{equation}
The bound in \eqref{eq:0nu2betaLLnumasslimit} can be translated into a limit on the amplitude
for \znbb decay at the parton level which can be estimated as
\begin{equation}
|\mathcal{A}_{0\nu\beta\beta}\up5 | \simeq
\frac{G_F^2}{ p_{\rm eff}^2 } 
|(m_\nu)_{ee}| \, ,
\label{eq:amplitude0nu2betaLL}
\end{equation}
where $p_{\rm eff} \sim 100$ MeV is the neutrino effective momentum 
obtained from averaging the corresponding nuclear matrix element contribution.
Thus, from \eqref{eq:0nu2betaLLnumasslimit} one obtains
\begin{equation}
\frac{p_{\rm eff}}{G^2_F}|\mathcal{A}_{0\nu\beta\beta}\up5 | \simeq
\frac{|(m_\nu)_{ee}|}{p_{\rm eff}} < 5\times 10^{-9}\, ,
\label{eq:amplitude0nu2betaLLlimit}
\end{equation}
which is the limit on the \znbb decay amplitude that we will also impose in the other two cases.
 
\subsubsection{LNV operators with one LH lepton and one RH charged lepton}

The leading operator of this class, given in Eq.~(\ref{eq:lowestoperatorLR}),
has dimension 7 because it must 
involve two leptons, three scalar doublets to cancel the leptonic hypercharge,
and one covariant derivative to compensate a chirality flip. Explicitly,
\beq
\ocal_{ea}\up{\oiv} = (\phid D^\mu \phit) \phid \overline{e_{eR}} \gamma_\mu
\lt_{aL} 
\rightarrow i \frac{g}{\sqrt{2}} v^3 W_\mu^-\overline{e_{eR}} \gamma^\mu
\nu^c_{aL}+\cdots \,.
\label{eq:IV7}
\eeq
It must be noted, however, that in order to write simple, working models
fulfilling 
all the experimental requirements, it may be necessary to impose
additional symmetries, and consequently, some of the external (light) scalar doublet
fields may not coincide with the SM Higgs doublet, as we mentioned above. 
This will be the case in the explicit model we will work out below.   

\begin{figure}
\begin{centering}
\includegraphics[width=0.55\columnwidth]{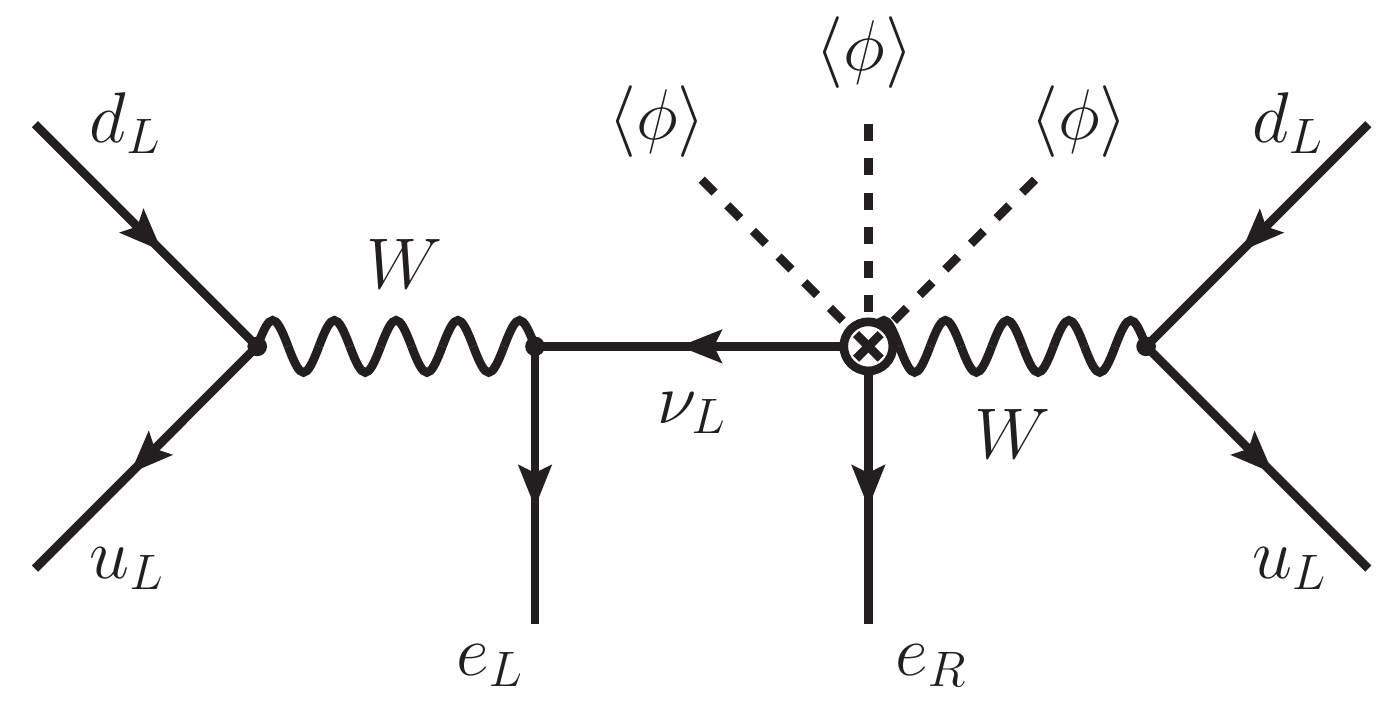}
\par\end{centering}
\caption{Tree-level diagram contributing to \znbb decay with 
one $\ocal\up7$ insertion.
\label{fig:0nu2betaLR}}
\end{figure}

The estimate from the \znbb decay amplitude shown in Fig. \ref{fig:0nu2betaLR}
is given by 
\beq
|\mathcal{A}_{0\nu\beta\beta}\up 7 | \simeq 
\frac{ G_F^2 v^3 |C_{ee}\up7|}{p_{\rm eff}\Lambda^3}\, .
\label{eq:amplitude0nu2betaLR}
\eeq
This translates into a bound on $\Lambda$, 
which must be
$> 100\, |C_{ee}\up7|^{1/3}\ \tev 
$, if we want $\mathcal{A}_{0\nu\beta\beta}\up7$ to satisfy the limit 
in \eqref{eq:amplitude0nu2betaLLlimit}.
These are order of magnitude estimates,  but one can also use detailed nuclear 
matrix elements available in the literature~\cite{Muto:1989cd,Pas:1999fc}. The interaction 
induced by the operator $\ocal\up7$
can be partially expressed as a modification of the standard weak interaction,
$W_\mu \overline{e}\gamma^\mu\left( (1-\gamma_5)+\eta (1+\gamma_5)\right)\nu$, 
where $\nu=\nu_L+\nu_L^c$ is a Majorana field. Then, the strong limit on $\eta$ 
derived using
detailed nuclear matrix elements calculations, $|\eta| < 4.4\times 10^{-9}$ 
(see \cite{Muto:1989cd} and \cite{Pas:1999fc} where $\eta$ was termed $\epsilon^{V+A}_{V-A}$),
reads in our case 
\begin{equation}
|\eta|=\frac{v^3}{\Lambda^3}|C\up7_{ee}| < 4.4\times 10^{-9}\, ,
\end{equation}
implying a bound which is very close to our estimate
\beq
\frac\Lambda{|C_{ee}\up7|^{1/3}} > 106\ \tev\, .
\label{eq:bound0nu2betaLR}
\eeq

\subsubsection{LNV operators with two RH charged leptons}

In this class the leading operator has dimension 9 and it is given in
 Eq.~(\ref{eq:lowestoperatorRR}):
\beq
\ocal_{ee}\up9 = \overline{e_{eR}} e_{eR}^c (\phid D^\mu \phit) (\phid D_\mu
\phit) 
\rightarrow -\frac{g^2}{2}v^4 W_\mu^- W^{-\mu}\overline{e_{eR}}
e_{eR}^c+\cdots\, .
\label{eq:9}
\eeq
\begin{figure}
\begin{centering}
\includegraphics[width=0.55\columnwidth]{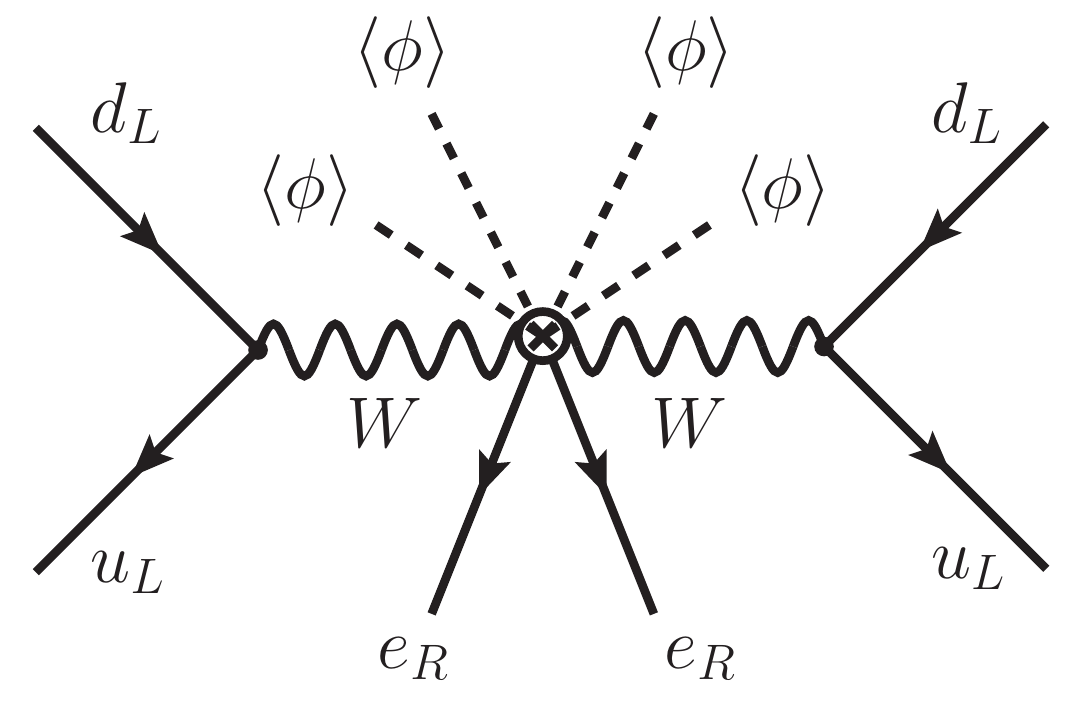}
\par\end{centering}
\caption{Tree-level diagram contributing to \znbb decay with 
one $\ocal\up9$ insertion.
\label{fig:0nu2betaRR}}
\end{figure}
As we have argued in the companion paper \cite{delAguila:2011gr}, 
the \znbb decay amplitude in Fig. \ref{fig:0nu2betaRR} can be large, 
near to its present experimental limit, with 
\begin{equation}
|\mathcal{A}_{0\nu\beta\beta}\up9 | \simeq 
\frac{ G_F^2 v^4 C\up9_{ee}}{\Lambda^5} \, . 
\label{eq:amplitude0nu2betaRR}
\end{equation}
Requiring that the $\mathcal{A}_{0\nu\beta\beta}\up9$  amplitude satisfies
the bound in Eq.~\eqref{eq:amplitude0nu2betaLLlimit}, one obtains 
$
\Lambda > 2\, |C_{ee}\up9|^{1/5}~\tev
$, 
but one can also make use of the detailed nuclear matrix element calculations. 
From $\ocal\up9$ one obtains the following 
six-fermion contact interaction inducing \znbb decay
\begin{equation}
\mathcal{L}_{0\nu\beta\beta}= 
\frac{G_{F}^{2}}{2m_{p}}\epsilon_3 \left(\bar{u}\gamma^{\mu}(1-\gamma_{5})d\right)
\left(\bar{u}\gamma_{\mu}(1-\gamma_{5})d\right)\bar{e}(1-\gamma_{5})e^{c}
\, ,
\label{eq:Lagrangian-0nu2beta}
\end{equation}
where $m_p$ denotes the proton mass and in our case
\begin{equation}
\epsilon_3=-\frac{2m_p v^4 C\up9_{ee}}{\Lambda^5}\, .
\label{eq:epsilon3}
\end{equation}
This type of interaction has been studied in \cite{Pas:2000vn}, where the bound
$|\epsilon_3| < 1.4\times 10^{-8}$ at 90\%~C.L. was found
\footnote{There is a misprint in Ref.~\cite{Pas:2000vn}. We thank the authors of this reference 
for providing us with the correct limit on $\epsilon_3$.}. This also leads to a limit very close to
our estimate
\begin{equation}
\frac\Lambda{|C_{ee}\up9|^{1/5}} > 2.7\ \tev\, .
\label{eq:bound0nu2betaRR}
\end{equation}
In Ref.~\cite{delAguila:2011gr} we present 
a realistic model which accommodates the observed neutrino masses and 
a large \znbb decay observable in the next round of  
experiments, proving explicitly the consistency of the previous estimates. 
In particular, neutrino masses are naturally suppressed even with NP 
at the TeV scale. 

These scale estimates are summarized in the Table~\ref{tab:t1}.
\begin{table}
\begin{tabular}{l|ccc}
TeV &\quad LL \quad & \quad LR \quad & \quad RR \quad \\
\hline
$\Lambda_{0\nu\beta\beta}$ & $10^{11}$ & $10^2$ & $1$ 
\end{tabular}
\label{estimates}
\caption{Natural (with $C_{ee}\up{n} \sim 1)$ NP scale limits for the three
different 
lowest order effective operators mediating \znbb decay.}
\label{tab:t1}
\end{table}
Note, however, that in actual models the operator coefficients
$C_{ee}\up{n} $ are not in general $\sim 1$ 
and the $\Lambda$ estimates may vary. For the LL case there are models 
with NP at $\Lambda \sim 1$ TeV for sufficiently suppressed  
couplings $C\up5 \sim 10^{-11}$ (see, for instance, Refs. \cite{Ma:2000cc,Aoki:2008av,Wei:2010ww,Ibarra:2010xw,Chen:2011de}). 
We can also have different scales within the model, as in the LR model 
below, where the new leptons can have masses below a TeV and 
be observable at the LHC. Similarly, in the RR case in 
Ref. \cite{delAguila:2011gr} the new scalar masses can range from 
few hundreds of GeV to tens of TeV. 
A more detailed discussion can be found in the companion paper.

\section{Majorana neutrino masses generated by the LNV operators inducing \znbb decay}
\label{sec:neutrinomasses}

Once the effective theory generates one of the LNV operators that produce \znbb decay, 
neutrinos will get a mass at some (loop) order, even if there is 
no other independent source of neutrino masses. 
The three operators not only stand for NP at quite 
different mass scales but result in 
different neutrino mass structures. Specific models
may of course include a combination of these effects,
so that the final structure may be quite involved.

\setcounter{subsubsection}{0}
\subsubsection{LL operator} 

After electroweak symmetry breaking $\ocal\up5$ generates a well-known
and much studied contribution to the neutrino masses (summation on repeated
indices must be 
understood when applicable through the manuscript):
\begin{equation} 
\frac{ C_{ab}\up5 }{\Lambda}\ocal\up5_{ab}= 
\frac{ C_{ab}\up5 }{\Lambda}\, \overline{\tilde{\ell}_{aL}}\phi \tilde \phi^\dagger
\ell_{bL}  
\rightarrow - \frac{v^2C_{ab}\up5}{\Lambda}  \, \overline{\nu_{aL}^c} \nu_{bL} \, , 
\label{Op5}  
\end{equation}
where $ a, b  = e,\mu,\tau $ are flavor indices.  Hence,
\beq
(m_\nu)_{ab} = \frac{2 v^2 {C\up5_{ab}}^*}{\Lambda} \, .
\label{eq:o5.mnu}
\eeq
In this case 
\znbb decay proceeds through the diagram in 
Fig. \ref{fig:0nu2betaSM}, and it is proportional to 
\begin{equation}
|(m_\nu)_{ee}| = | U_{ei} m_i U^{T}_{ie}  | 
= | c^2_{13} c^2_{12} e^{i\alpha_1} m_1 + c^2_{13} s^2_{12} e^{i\alpha_2} m_2 
+ s^2_{13} e^{-2i\delta} m_3 | \; , 
\label{Effelectronneutrinomass}
\end{equation}
where $U$ is the mixing matrix diagonalizing the neutrino 
mass matrix written in the current eigenstate basis with 
well-defined charged lepton flavor \cite{Pontecorvo:1967fh,Maki:1962mu}: 
\begin{equation}
m_{\nu}=\frac{2  v^2 {C\up5}^\dagger}{\Lambda} =U\left(\begin{array}{ccc}
                                                       m_1 &  & \\
                                                         & m_2 & \\
                                                        &  & m_3
                                                      \end{array} \right)U^{T}
\label{masses}
\end{equation}
and
\begin{eqnarray}
U=\left(\begin{array}{ccc}
c_{13}c_{12} & c_{13}s_{12} & s_{13}e^{-i\delta}\\
-c_{23}s_{12}-s_{23}s_{13}c_{12}e^{i\delta} &
c_{23}c_{12}-s_{23}s_{13}s_{12}e^{i\delta} & s_{23}c_{13}\\
s_{23}s_{12}-c_{23}s_{13}c_{12}e^{i\delta} &
-s_{23}c_{12}-c_{23}s_{13}s_{12}e^{i\delta} & c_{23}c_{13}
\end{array}\right)\left(\begin{array}{ccc}
e^{i\alpha_{1}/2}\\
 & e^{i\alpha_{2}/2}\\
 &  & 1
\end{array}\right)\;,
\label{UPMNS}
\end{eqnarray} 
where $s_{ij} \equiv \sin\theta_{ij}$ and $c_{ij} \equiv \cos\theta_{ij}$. 
Thus $\Lambda$ is required to be $\sim 6 \times 10^{11}$ TeV to reproduce 
the observed neutrino masses $(m_\nu)_{\tau\tau} \sim 0.1$ eV for $C_{ab}\up5 \sim 1$,
which is of the same order as the limit derived from \znbb decay (see
Table \ref{tab:t1}).

\subsubsection{LR operator} 

Restoring flavor indices the LR contribution to the effective Lagrangian
becomes
\begin{equation} 
\frac{C\up7_{ab}}{\Lambda^3} (\phid D^\mu \phit) 
(\phid \overline{e_{aR}} \gamma_\mu \lt_{bL})\; . 
\label{L7}
\end{equation} 
In this case the neutrino masses are generated by radiative
corrections (left graph in Fig.~\ref{fig:1L2Lmasses}).
Here it is important to differentiate between the
calculable (logarithmic) contributions to the masses derived from $ \ocal\up7$
and the estimates obtained by matching. The first, obtained in
standard effective field-theory fashion using dimensional regularization
and a renormalizable gauge, scale like
\beq
(\delta m_\nu)_{ab} \simeq \frac{v^3}{16 \pi^2 \Lambda^3} \left(m_a C\up7_{ab} + 
m_b C\up7_{ba}\right)\log\left(\frac{\Lambda}{v}\right)\, .
\eeq
Whereas the estimates from matching, which give the dominant contribution to neutrino 
masses, are obtained using dimensional analysis and are of the form
\beq
(m_\nu)_{ab} \sim \frac{v}{16 \pi^2 \Lambda} \left(m_a C\up7_{ab} + 
m_b C\up7_{ba}\right)\, ,
\label{eq:o7mnu}
\eeq
as we will derive for the specific models discussed below.
Note, however, that even for this last estimate it is important
to use a renormalizable gauge as in the unitary gauge spurious
positive powers of $\Lambda$ may appear. That the final result must
be proportional to $1/\Lambda$ follows from {\it(i)} the fact that the
dimension 5 Weinberg operator is the only one describing
neutrino Majorana masses (in the absence of $\nu_R$); and {\it(ii)} that
we assume that the NP is decoupling.
It is also important to note that, as mentioned in the
introduction, specific models may have
more than one NP scale so that $ \Lambda $ represents an
effective scale and  may not correspond to the mass of any 
specific particle. For example, if the theory has two scales
$M>M'$ one may have $ \Lambda = M^2/M' $; we will present
an example of this situation in Section \ref{sec:LR-model}.

The expression (\ref{eq:o7mnu})
coincides with the results obtained in specific models such as the
one worked-out in  Section \ref{sec:LR-model}
(Eq. (\ref{modelneutrinomass})) with the appropriate identification of
$\Lambda $.
In this case $(m_\nu)_{\tau\tau} \sim 0.1$ eV with $C\up7_{\tau\tau} \sim 1$
implies 
(see Eq. (\ref{al:massesLRcorrect})) 
$\Lambda \sim 4\times 10^7$ TeV, that should be compared to the limit
obtained from \znbb decay in Table \ref{tab:t1}, which is several orders of magnitude
smaller. This implies that in realistic models the coefficients $C\up7$ must be
much less than $1$ to allow for a $\Lambda$ of the order of $100$~TeV, in agreement with the 
$0\nu\beta\beta$ decay estimate, if this is to be observed in the next generation of 
experiments. This is what happens in the explicit model we will work out below.

\subsubsection{RR operator} 

Finally, the LNV operator $\ocal\up9$ generates Majorana 
masses to neutrinos at two loops, in this case also suppressed 
but by two loop factors and two charged lepton mass 
insertions (see the right panel of Fig.~\ref{fig:1L2Lmasses}).
As for the LR case we distinguish between the calculable (logarithmic)
contributions,
\beq
(\delta m_\nu)_{ab} \simeq \frac{v^4}{(16 \pi^2)^2 \Lambda^5} m_a C\up9_{ab} m_b 
\log\left(\frac{\Lambda}{v}\right)\, , 
\eeq
and the  estimates obtained from matching
\begin{equation} 
(m_\nu)_{ab} \sim \inv{(16 \pi^2)^2 \Lambda} m_a C\up9_{ab} m_b \, . 
\label{eq:massesRR}  
\end{equation} 
Explicit calculations
in specific models (using renormalizable gauges) \cite{delAguila:2011gr}
reproduce \eqref{eq:massesRR} up to a proportionality factor of order 1
that depends on the various masses in the loop, and when $ \Lambda $
is identified with an appropriate combination of heavy scales.
For $(m_\nu)_{\tau\tau} \sim 0.1$ eV and 
$C\up9_{\tau\tau} \sim 1$, \eqref{eq:massesRR} implies
$ \Lambda \sim 1.3 \times 10^3\ \tev $, several orders of magnitude larger than 
the \znbb decay estimate in Table \ref{tab:t1}. Although in realistic models 
the coefficients $C\up9$ and $\Lambda$ are typically
smaller~\cite{delAguila:2011gr}.

\section{Heavy particle additions generating the lowest order LNV operators at
tree level}
\label{sec:mediators}

In this section we will work out the combinations of heavy particles
(scalars, fermions and gauge bosons) that must be present in any
extension of the SM if it is to generate one of the operators
$ \ocal\up{5,7,9}$ at tree level; thus producing the largest possible rates for 
\znbb decay. We will assume that the underlying theory is weakly coupled
and contains only renormalizable vertices~\footnote{Non-renormalizable
vertices are presumably suppressed by inverse powers of a yet higher scale
$ \Lambda_{\rm high} $ that we assume much larger than $ \Lambda $.};
we also assume the NP respects all the gauge symmetries of the SM. 
In listing the heavy particles we denote by
$\vec IY,~ \fer IY $ and $ \sc IY $
a heavy vector, fermion or scalar with isospin
$I$ and hypercharge $Y$, respectively. 
When the heavy particles can be either a heavy
vector or heavy scalar with the same
isospin and hypercharge, we use $ \bos IY$
to denote both possibilities.

\setcounter{subsubsection}{0}
\subsubsection{LL additions}

\begin{figure}
\begin{centering}
\includegraphics[width=0.95\columnwidth]{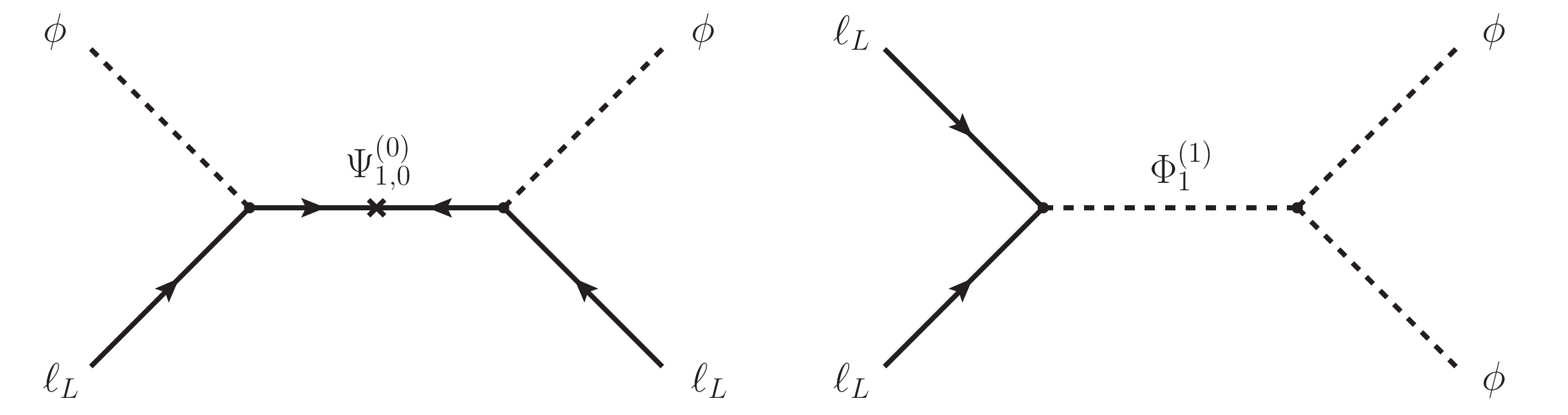}
\par\end{centering}
\caption{Topologies generating $\ocal\up5$.
These graphs together
with those with one or two $W$ boson attachments
generate $\ocal\up{\oi,\oii,\oiii}$.
\label{fig:topologiesII5}}
\end{figure}

In the case of {\it LNV operators with two LH leptons} the heavy excitations
that can 
generate the operators of dimension 5 and 7 in Eqs. (\ref{eq:operatorll5}) and 
(\ref{eq:operatorsll7}), respectively, are 
\bea
\sc{1}1\xor \fer{1,0}0\; . 
\label{O5mediators}
\eea
(If the model has two or more light scalar isodoublets, a heavy scalar singlet 
$\sc{0}1$ can also generate these operators~\cite{Oliver:2001eg}.) 
This means that, as mentioned previously,
if the underlying theory generates (\ref{eq:operatorsll7}), it will also
generate (\ref{eq:operatorll5}) with the {\em same} heavy scale. 
In Fig. \ref{fig:topologiesII5} we depict the diagram topologies resulting in 
those operators upon integration of the heavy particles flowing 
through the internal lines. 

\subsubsection{LR additions}

The sets of heavy excitations that can generate the dimension 7 
{\it LNV operator with one LH lepton and one RH electron} 
in Eq. (\ref{eq:IV7}) are 
\bea
\{\vec{1/2}{3/2}, \bos{0,1}1 \}
 \xor 
\{\fer{1/2}{1/2}, \bos{0,1}1 \}
  \xor 
\{\fer{1/2}{1/2}, \fer{0,1}0 \}
\xor 
\{\fer00, \bos01 \}
 \xor 
\{\fer10, \bos11 \} \,.
\label{eq:IV7H}
\eea
In this case one must always exchange two heavy particles. 
It is worth noting that in the last two possibilities the
heavy fermions $\fer{0,1}0$ necessarily have couplings
that would also generate the operators in Eqs. (\ref{eq:operatorll5}) 
and (\ref{eq:operatorsll7}) at tree level. In contrast, models containing
$\sc{0,1}1$ may or may not generate them, 
depending on whether the heavy scalars couple to $ \ell_L \times \ell_L $ 
(which may be forbidden by the symmetries of the underlying theory). 
In the appendix we provide the diagram topologies relevant to 
each case. 

One can construct many models choosing from the above NP matter contents. 
However, in order to make the model phenomenologically viable we must often
enlarge these 
minimal sets. This is because (\ref{eq:IV7H}) are fixed only by the requirement
that they generate $ \ocal\up7 $ at tree level, which does not insure
the preservation of extra 
symmetries that are sometimes necessary, for instance, to avoid too large LFV
rates, or 
to forbid tree-level neutrino masses, or to reproduce the observed lepton
spectrum.
In Section \ref{sec:models} we discuss a realistic, simple example for the case 
$\{\fer{1/2}{1/2}, \sc11 \}$. But, as we will argue, for the model to be
realistic, it 
must include at least two 
fermion doublets $\fer{1/2}{1/2}$ besides one scalar triplet $\sc{1}1$,
and, in addition, a second light scalar doublet $\phi'$. 

\subsubsection{RR additions}

Finally, the sets of heavy excitations that can generate at tree level the
dimension 9 
{\it LNV operator with two RH electrons} in Eq. (\ref{eq:9}) are 
\bea
&\{\sc02, \bos{1/2}{3/2}, \bos{0,1}1 \}
 \xor 
\{\sc02, \bos{0,1}1 \} 
\xor 
\{\fer{1/2}{1/2}, \bos{1/2}{3/2}, \bos{0,1}1 \}
 \xor \cr
&\{\fer{1/2}{1/2}, \fer10, \bos{0,1}1 \} 
  \xor 
\{\fer{1/2}{1/2}, \bos{0,1}1 \} 
 \xor 
\{\fer10, \bos11 \}
 \xor 
\{\fer00, \bos01  \}\;.
\label{eq:O9.NP}
\eea
Despite the presence of $\sc{0,1}1 $ and $ \fer{0,1}0$ in some of these
options, these heavy particles need not have the same vertices as the ones
leading to (\ref{eq:operatorll5}). If they do, $ \ocal\up9 $ would
have only subdominant effects; but this is in general not the case.
In the appendix we also provide the diagram topologies relevant to 
each of these cases.

There are many models that can be constructed containing the above
particle content. For example, the case where the scalar sector of the SM
is extended 
by adding a doubly-charged isosinglet $\sc02$ and an isotriplet of
unit hypercharge $ \sc11 $, was considered in detail
in the companion paper \cite{delAguila:2011gr}. 
(See also \cite{Chen:2006vn,Chen:2007dc}.) 
As in other cases, additional heavy fields may be required 
in order to make the model realistic. 

\section{Simple examples of fundamental theories with a large \znbb decay rate 
and realistic neutrino masses}
\label{sec:models}

There are many specific models fulfilling the three generic scenarios discussed 
above (see Eqs. \eqref{eq:lowestoperatorLL},\eqref{eq:lowestoperatorLR} 
and \eqref{eq:lowestoperatorRR}). Here we list a few for illustration purposes;
we do not aim at reviewing
all models that have been considered in the literature; concentrating instead
on specific examples that exhibit the salient features discussed previously.
In particular neutrino masses take the form in Eqs.~\eqref{al:massesLLcorrect},
\eqref{al:massesLRcorrect} and \eqref{al:massesRRcorrect}, and in general the coefficients 
$C\up{n}\ll 1$
for consistency in realistic models.

\setcounter{subsubsection}{0}
\subsubsection{LL models} 

SM extensions that at tree level generate the Weinberg operator $\ocal\up5$ 
at low energies, have been extensively
studied in the literature  (for type I see-saw
see~\cite{Minkowski:1977sc,GellMann:1980vs,Yanagida:1979as,Mohapatra:1979ia},
for see-saw type II
see~\cite{Konetschny:1977bn,Cheng:1980qt,Lazarides:1980nt,Magg:1980ut,
Schechter:1980gr,Mohapatra:1980yp},
and for see-saw type III see \cite{Foot:1988aq,Ma:2002pf}). Specific models for
any of the three possible see-saw 
scenarios~\footnote{Assuming only one light SM Higgs doublet.}
(see Eq. (\ref{O5mediators})) require very heavy mediators, 
with masses $\sim 10^{14}$ GeV or very small couplings,  and
were discussed some time ago (for recent reviews 
see~\cite{King:2003jb,Altarelli:2004za,Mohapatra:2005wg,GonzalezGarcia:2007ib}
). 
The most popular and simplest case, and also the pioneering one, 
results from the addition of heavy RH neutrinos $\fer{0}0 = \nu_R$ with 
the renormalizable Lagrangian 
\begin{equation}
\mathcal{L}^{\nu_R} = i\, \overline{\nu_{aR}}\, \slashed{\partial}\nu_{aR} -
\{\frac{1}{2} M_a\overline{\nu_{aR}^c}\, \nu_{aR} -
y_{ab} \overline{\nu_{aR}}\, {\tilde\phi}^\dagger \ell_{bL} 
+ \mathrm{h.c.} \} \ ,
\label{Heavyneutrinos}
\end{equation}
including the kinetic terms and the Yukawa couplings $y_{ab}$. 
There must be at least two heavy neutrinos $\nu_{aR}$ (with masses $M_a$) 
to guarantee that the light neutrino mass matrix is at least of rank 2,
required in order to account for the three non-degenerate light neutrinos. Explicitly,
\begin{equation}
(m_\nu)_{ab} = \frac{2v^2 C^{(5)*}_{ab}}{\Lambda}=- \frac{y^*_{ca}y^*_{cb}}{M_c} v^2 \ , \mathrm{with}\ 
\left| \frac{C\up5_{ab}}{\Lambda}\right| \lesssim
10^{-11}\ \tev^{-1}\, ,
\label{O5neutrinomasses}
\end{equation}
in agreement with Eq.~\eqref{eq:LambdaC5ee} and Table~\ref{tab:t1}. 
The alternative case with the see-saw messengers near the TeV scale (and
$|y|<10^{-5}$) 
has become more popular with the launch of the LHC, see
\cite{Ibarra:2010xw,Ibarra:2011xn}, and references therein. 

In this scenario \znbb decay follows from the exchange of the 
light Majorana neutrinos; the resulting amplitude is proportional to the 
effective electron-neutrino mass $(m_\nu)_{ee}$ in Eq.~(\ref{Effelectronneutrinomass}), 
see Fig.~\ref{fig:0nu2betaSM}; hence, any \znbb decay rate  
within the present experimental precision can be accommodated. 
Indeed, a global fit to neutrino oscillation data gives (see, for instance,
\cite{Schwetz:2011zk})
$\Delta m_{21}^{2} \equiv
m_{2}^{2}-m_{1}^{2}=(7.59^{+0.20}_{-0.18})\times10^{-5}\:\textrm{eV}^{2}$,
$\Delta m_{31}^{2} \equiv
m_{3}^{2}-m_{1}^{2}=(2.50^{+0.09}_{-0.16})\times10^{-3}\:\textrm{eV}^{2}$,
$s_{12}^{2}=0.312^{+0.017}_{ -0.015}$, $s_{23}^{2}=0.52^{+0.06}_{ -0.07}$, 
$s_{13}^{2}=0.013^{+0.007}_{ -0.005}$. Replacing these values into 
Eq.~(\ref{Effelectronneutrinomass}) one can obtain at $1 \sigma$ any \znbb decay 
rate compatible with present experimental limits for the normal hierarchy, 
although it is bounded from below 
for the inverse one \cite{Nakamura:2010zzi,Bilenky:1999wz,Bilenky:2001rz}.  

\subsubsection{LR models} 
\label{sec:LR-model}

To our knowledge, no realization of this second scenario has been spelled out
in the literature. 
One can construct many models with \znbb decay into two 
electrons of opposite chirality mediated by $\ocal\up{\oiv}$ by choosing 
among the matter contents in Eq. (\ref{eq:IV7H}); 
however, when constructing a realistic  model we must in general enlarge 
these minimal sets.

\begin{figure}
\begin{centering}
\includegraphics[width=0.55\columnwidth]{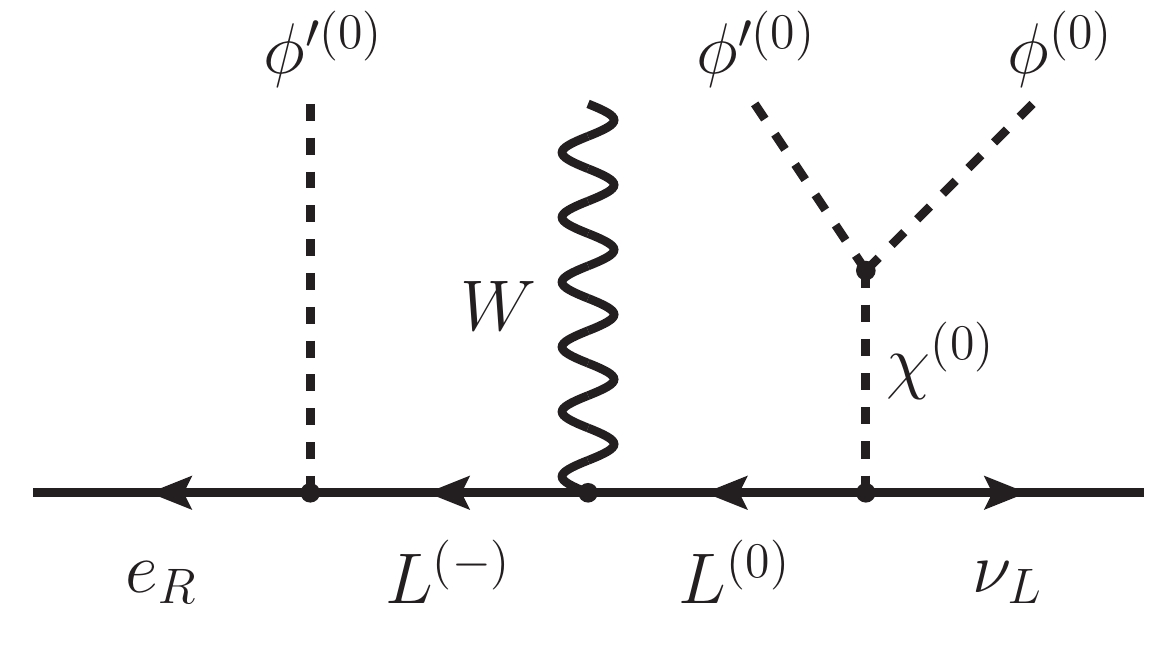}
\par\end{centering}
\caption{Tree-level diagram contributing to $\ocal\up{\oiv}$ in the model
proposed.
\label{fig:IV7generation}}
\end{figure}

We start from the set $\{\fer{1/2}{1/2}, \sc11 \}$ contained in Eq.
(\ref{eq:IV7H});
to simplify the notation we define  $\sc{1}1 \equiv  \chi$,
a scalar isotriplet  of hypercharge $1$, and $\fer{1/2}{1/2} \equiv L^c = L^c_L
+ L^c_R$,
a lepton isodoublet  of hypercharge $1/2$
(in terms of its LH and RH components); a
simple way to insure the 
decoupling of the heavy physics is to assume, as we do,
that the heavy fermions are vector like. 
This particle content
is sufficient to generate
$ \ocal\up7 $ at tree level, and it is not hard to convince oneself
(see the appendix) that the relevant graphs must involve
the couplings $e_R\phi \tilde{L}$, $ \ell_L L \chi$ and $
\phi^\dagger \phi^\dagger \chi $. However, such a model
also allows the coupling $\ell_L \ell_L  \chi $ and will
then generate $ \ocal\up5 $ at tree level through the
standard type-II see-saw diagram (on the right of Fig.~\ref{fig:topologiesII5}),
a possibility we wish to disallow.
In order to do this we impose a discrete $Z_2$ symmetry under which
$ \chi$ and $L$ are odd and $ \ell_L $ is even, so $ \ocal\up5$
does not appear at tree level; unfortunately, this symmetry
also forbids the $\phi^\dagger \phi^\dagger \chi $ vertex . In
order to overcome this difficulty we assume the presence of two
light scalar doublets $ \phi,~\phi' $ which are, respectively, 
even and odd under $Z_2 $. The allowed vertices are then
\beq
\overline{\tilde{L}} \chi \ell_L\, , \qquad
\overline L \phi' e_R\, , \qquad
\phi^\dagger \chi \tilde{\phi'}\ , 
\eeq
from which $ \ocal\up7$ is generated
through diagrams such as the one in Fig. \ref{fig:IV7generation},
while $ \ocal\up5 $ appears only at one loop.

As we will explain below, in order to accommodate
a generic neutrino mass matrix and also 
allow for flavor symmetries treating the three families on the same footing,
we will assume the presence of 3 heavy vector-like fermion doublets $L_a,~
a=1,2,3$.  
The complete list of new fields is given in Table \ref{tab:modelfields};
the Lagrangian will include renormalizable couplings preserving these symmetries, 
noting that 
the SM fields transform trivially under $Z_2$.  
The second scalar doublet $\phi^\prime$ 
could be identified with the isodoublet giving mass to the up 
quark sector in two-doublet
models~\cite{Lee:1973iz,Weinberg:1977ma,Wilczek:1977pj,Deshpande:1977rw,
Haber:1978jt,Donoghue:1978cj}
(for reviews see for instance \cite{Gunion:1989we,Branco:2011iw}) 
if we require the RH up-quark singlets to be odd under $Z_2$
(the quark isodoublets are even). We will prefer not to do so, 
because we will find it convenient to be able to assign a
small VEV $\vevof{\phi'}$.
We also note that in this model
LN is explicitly broken by (renormalizable) terms in the scalar potential, 
in particular by $(\phi^\dagger  \phi^\prime)^2$. 
\begin{table}
\begin{tabular}{l|cccc}
$ $ &\quad $L_{L a}$ \quad & \quad $L_{R a}$ \quad & \quad $\chi$ \quad 
& \quad $\phi^\prime$ \quad \\
\hline
$SU(2)_L$ & $\frac{1}{2}$ & $\frac{1}{2}$ & 1 & $\frac{1}{2}$ \\
$U(1)_Y$ & $-\frac{1}{2}$ & $-\frac{1}{2}$ & 1 & $\frac{1}{2}$ \\
$Z_2$ & $-$ & $-$ & $-$ & $-$  
\end{tabular}
\caption{Quantum number assignment for the extra fields.\label{tab:modelfields}}
\end{table}

The scalar potential can be easily arranged to insure a  
minimum where  $\vevof{\phi} \gg\  \vevof{\phi^{\prime}}, 
\vevof{\chi}\  \neq \ 0$, 
with $\vevof{\chi} \simeq -\mu^* 
\vevof{\phi^{\prime}} \vevof{ \phi} /m^2_\chi$,
$\mu$ the trilinear $\phi^\dagger \chi \tilde{\phi'}$
coupling
and $m_\chi$ the isotriplet mass
(in order to satisfy the limit from electroweak precision 
data \cite{Nakamura:2010zzi,delAguila:2008ks}
we require $ \vevof{\chi} \lesim 2$ GeV;
see \cite{delAguila:2011gr} for a similar analysis and \cite{Kanemura:2012rs} for a recent study, in the framework of
the type II see-saw, including one-loop radiative corrections from the scalars of the model).
We  assume negative mass terms for $\phi$ and $\phi^\prime$ to trigger the 
corresponding VEVs, 
whereas $\chi$ gets a VEV through its mixing with the scalar isodoublets. 
Otherwise, dimensional couplings in the potential are typically of electroweak
order, 
except for new scalar masses that may be larger. 
Dimensionless ones stay perturbative, in general ranging within 
an $\alpha_{EM} \sim 10^{-2}$ factor. 

We now discuss briefly the viability of the model, concentrating on
the effects of the masses and mixings of the new fermions, 
the induced LFV effects, and the implications for both the 
LHC, and the light neutrino masses.  The heavy lepton Lagrangian reads 
\begin{equation}
\mathcal{L}^L_{\mathrm{H}}= \overline{L_{a}} (i \cancel{D} - M_a) L_a 
+ \{y^e_{ab} \overline{L_{aL}} \phi^\prime e_{bR} 
+ y^\nu_{ab} \overline{\tilde{L}_{aL}} \chi \ell_{bL} 
+ \mathrm{h.c.} \} \ ,
\label{Heavy}
\end{equation}
where we assumed that the heavy mass matrix is diagonal 
without loss of generality. Once $ \phi' $ and $\chi$
acquire VEVs the light $e_R$, $\ell_L$ leptons
mix with the $L_a$; such mixings and the
corresponding phenomenology of 
heavy vector-like lepton doublets were analysed
long ago \cite{delAguila:1982yu,Langacker:1988ur,Nardi:1991rg}, 
and  more  recently within the context of 
Little Higgs models
 \cite{Blanke:2007db,delAguila:2008zu,delAguila:2010nv,Goto:2010sn}, 
and Extra Dimensional theories \cite{Huber:2001ug,Carena:2009yt,delAguila:2010vg}  
(for a review see \cite{Raidal:2008jk,Feldmann:2011zh}; for updated limits 
see \cite{delAguila:2008pw}).

The low-energy effects of such
mixings will be proportional to $ y^e_{ab} \vevof{\phi'}/M_a$
or $ y^\nu_{ab} \vevof{\chi}/M_a$ and
can be made as small as experimentally required
by increasing the heavy masses 
$M_a$, reducing the couplings $y^{e,\nu}_{ab}$, or
the VEVs $\vevof{\phi^\prime}, \vevof\chi$.
LFV effects can be further suppressed by
assuming that the light charged leptons, which get their masses through 
the SM Higgs mechanism, are aligned along the heavy flavors. This 
corresponds to taking $y^e_{ab}$ diagonal, which may be natural in a larger
model. The LHC reach for the scalar triplet was reviewed in
the companion paper and is updated in next section.
The production of the $L_a$ at the LHC
has been also studied previously \cite{delAguila:1989rq} 
and the general conclusion is that they will be detected
provided their masses are below $850$~GeV for a center of mass (CM) energy
of $14$~TeV and an integrated luminosity of $100$~fb$^{-1}$ \cite{AguilarSaavedra:2009ik}
(the LHC reach reduces to $350$~GeV for heavy leptons
mainly decaying into taus \cite{delAguila:2010es}).

Given the couplings of the model one can evaluate $C\up7_{ab}$ by using the diagram
in Fig. \ref{fig:IV7generation},
\begin{equation}
\frac{C\up7_{ab}}{\Lambda^3} = -i\frac{\mu y^{e*}_{ca}y^{\nu*}_{cb}}{m^2_\chi M^2_{c}}\, , 
\label{eq:C7model}
\end{equation}
where all masses in the $L_c$ and $\chi$ multiplets are taken equal.

\begin{figure}
\begin{centering}
\includegraphics[width=0.55\columnwidth]{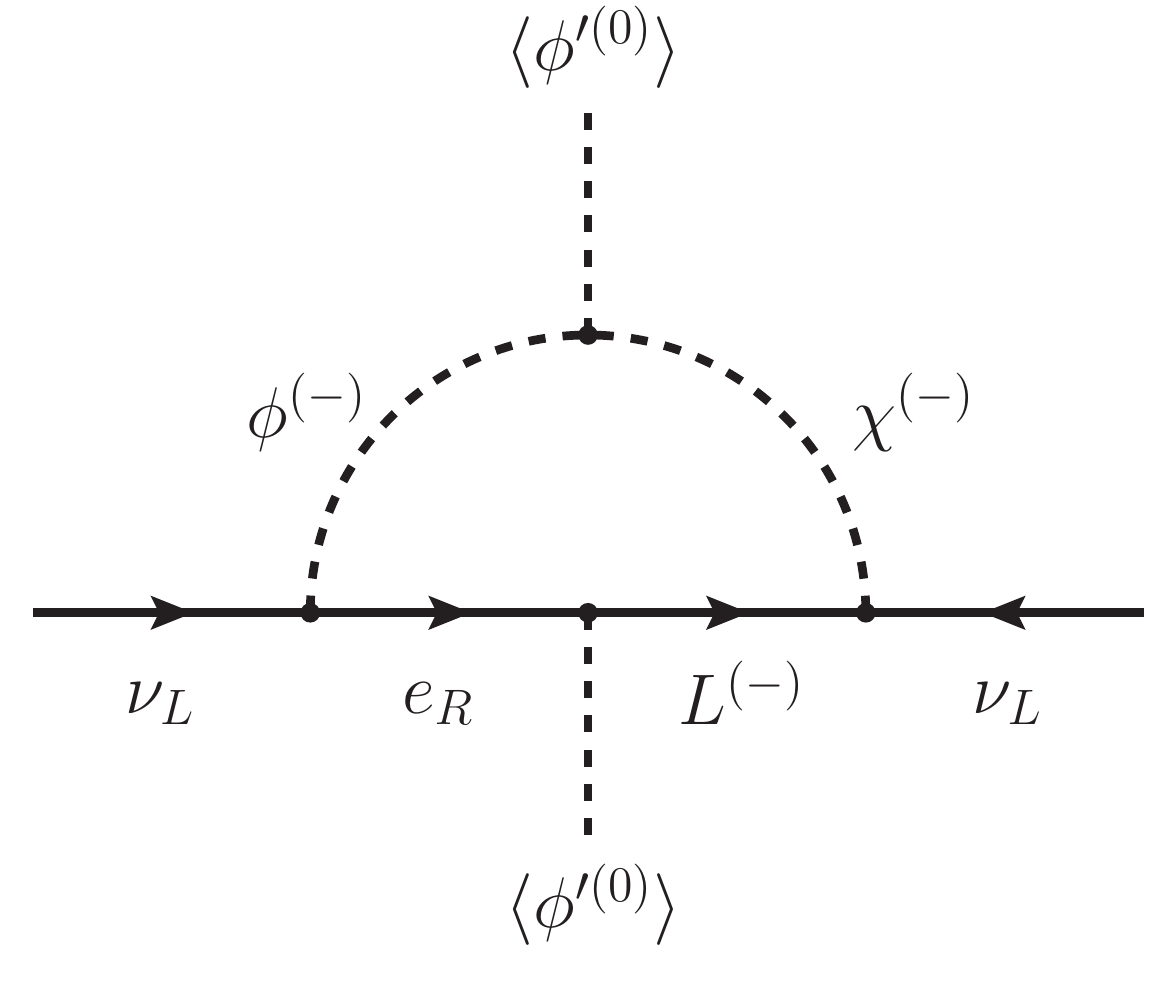}
\par\end{centering}
\caption{Leading one-loop contribution in the Feynman gauge to 
neutrino masses in the fundamental theory.
\label{fig:neutrinomassesLRmodel}}
\end{figure}

Light neutrinos are massless at tree level, but 
they get a mass at one loop through diagrams like that in 
Fig. \ref{fig:neutrinomassesLRmodel}. 
The full calculation gives 
\begin{equation}
(m_\nu)_{ab} \simeq \frac{v^{\prime\, 2}\mu}{32 \pi^2 v} 
\left(m_a y^{e *}_{ca} y^{\nu*}_{cb} + m_b y^{e *}_{cb} y^{\nu*}_{ca}\right) 
\frac{1}{M^2_c-m_\chi^2}\log \frac{M^2_c}{m^2_\chi}
\; , 
\label{modelneutrinomass}
\end{equation}
where  $v^\prime=\vevof{\phi^{\prime}}$
and we have assumed that all other masses are much smaller than 
$M_c$ and $m_\chi$.
Thus, with only one heavy lepton doublet the neutrino mass matrix 
has at most rank 2. With 
two heavy lepton doublets all three light neutrinos can be massive, but 
with 
three it is also straightforward to impose a flavor symmetry that
forbids any  potentially large LFV. 

The model may face a domain wall problem
if the spontaneously broken $Z_2$ symmetry is exact. This can be obviated adding
a softly breaking term $\phi^\dagger \phi^\prime$ to the scalar potential.
In a future publication we will provide a detailed analysis of this model 
including the neutrino 
mass calculation, as well as the quantitative discussion of the parameter space 
allowed by present experimental bounds, and the predictions for the 
different observables. 

\subsubsection{RR models} 

The final case, where \znbb decay involves two RH electrons
through the operator $\ocal\up9$, allows for many tree-level realizations, 
typically with a heavy sector near the electroweak scale as argued in 
Section \ref{sec:mediators} and in the appendix.
A realistic, simple model of this scenario with the neutrino masses generated
at 
two loops, and predicting a non-zero third mixing 
angle $\sin^2 \theta_{13} \gtrsim 0.008$ for a large \znbb decay rate, 
was studied thoroughly in the companion paper 
\cite{delAguila:2011gr} and corresponds to the  particular 
SM addition $\{\sc02, \sc11 \}$ in Eq. (\ref{eq:O9.NP}). 
The model contains in addition
a real scalar singlet $\sigma$ whose presence allows 
a $Z_2$ discrete symmetry that protects the 
neutrinos from acquiring a tree-level mass. This, in turn, makes the neutrino 
masses calculable.

\section{Other phenomenological implications}
\label{sec:phenomenology}

Although it is beyond the scope of this paper to make 
a detailed discussion of the experimental constraints 
on the different scenarios and models, we will briefly 
comment on the main phenomenological implications 
for ongoing and forthcoming experiments. 
The present study answers the general question of 
how large can the \znbb amplitude be, requiring 
that the only light fields are those of the SM, and 
assuming that the NP responsible for the effect is
weakly coupled and has a characteristic scale above
the electroweak scale. 
In particular this means then that neutrino masses are 
Majorana; we also assume, as it is widely believed, 
that the observed pattern of neutrino oscillations 
indicates the presence of three massive neutrinos~\cite{Nakamura:2010zzi}. 
This strongly influences our analysis because we must then 
explain the neutrino spectrum, assuming that there is no other 
larger source of LNV beyond the one mediating 
\znbb decay.

Thus, within this framework, we have to verify on a case 
by case basis whether:  
\begin{itemize}
\item
The \znbb decay rate can be fast enough to be 
observable at the next round of experiments. 
\item
The neutrino masses are correctly predicted.
\item
LFV and universality limits are within experimental bounds. 
\item
The new heavy particles satisfy the collider exclusion limits.  
\end{itemize}
We have already addressed the first two points in general and in 
some detail; we consider the remaining two
in the following subsections.

\subsection{LFV processes in SM extensions with a sizeable \znbb decay rate}

A relatively large \znbb decay rate requires a not too heavy 
NP (except for the case where $\ocal\up5$ is generated at tree level). 
This, in turn, implies that the effects of the
new particles may be detected in highly 
suppressed processes, like in flavor-changing
leptonic or $Z$ decays \cite{Nakamura:2010zzi};
the corresponding constraints are model dependent, however. 
We can compare, for illustration, the LR and RR models in the 
previous section and in Ref. \cite{delAguila:2011gr}, respectively. 
In the latter the heavy sector only involves new scalars, with little 
effect on lepton universality precision tests, for instance. 
The main universal constraint derived from them being the upper bound 
on the isotriplet VEV, $\vevof{\chi} \lesim 2$ GeV 
\cite{Nakamura:2010zzi,delAguila:2008ks}. 
This limit is easy to satisfy although it is rather restrictive in 
this specific model due to the small number of new free parameters 
available, and the desirability of accommodating a large \znbb decay rate.
For the same reason LFV constraints are quite demanding 
because the corresponding leptonic decays are proportional 
to neutrino masses, which have a large misalignment from the charged 
lepton current eigenstates,  described by the Pontecorvo-Maki-Nakagawa-Sakata 
mixing matrix \cite{Pontecorvo:1967fh,Maki:1962mu}. 
This translates into lower bounds on the heavy scalar masses 
with a preference for values above a TeV \cite{delAguila:2011gr}. 

The LR model presented in this manuscript is somewhat different, 
for the new scalars have no direct coupling to SM lepton pairs, 
although they are induced by fermion mixing. 
In this case the most stringent restrictions concern the new heavy 
leptons, with masses $M_a$, especially on their mixing with the light ones;
such indirect constraints on extra vector-like leptons have been 
thoroughly studied  in the literature
\cite{delAguila:1982yu,Langacker:1988ur,Nardi:1991rg,
Blanke:2007db,delAguila:2008zu,delAguila:2010nv,Goto:2010sn,
Huber:2001ug,Carena:2009yt,delAguila:2010vg,
Raidal:2008jk,Feldmann:2011zh,delAguila:2008pw}.
The vector-like character that allows their decoupling without breaking the
SM, 
also fixes their mixing behavior and  their low energy phenomenology. 
The corrections to SM vertices are suppressed by at least 2 
powers of a small ratio ${\cal O} (m_\ell/M_a)$, where $m_\ell$ 
is typically a light lepton mass. Hence, the decay rates, which 
are vanishingly small within the SM, are suppressed by at 
least 4 powers of these ratios; moreover, they also vanish 
when $\vevof{\phi'}\ \rightarrow 0$. 
Even more, LFV processes can be also canceled by aligning 
the heavy and light lepton flavors. In summary, many different 
small factors can conspire to make negligible the effects of
 the heavy fermion, in particular, their tree-level effects
are naturally small enough to accommodate the experimental constraints. 
The main restrictions on this type of models result from one-loop 
contributions exchanging heavy leptons and bosons; 
the most restrictive processes being those including 
the muon to electron transition. In fact, one can, to a large extent, 
apply the conclusions from related analyses 
for the Littlest Higgs model with T-parity
\cite{Blanke:2007db,delAguila:2008zu,Goto:2010sn,delAguila:2010nv};
the general conclusion is that the
heavy flavors must be aligned with the light charged leptons 
with a precision better than $1-10$ \% for heavy masses 
of ${\cal O}({\rm TeV})$. 

In summary, LFV provide stringent restrictions on these 
models but they can be satisfied within relatively large regions 
of parameter space. What would be more interesting, 
these models could also explain a departure from the 
SM predictions if found in the ongoing and forthcoming 
searches for LFV~\cite{Nakamura:2010zzi,Adam:2011ch} (see \cite{Cei:2010zz} for a review).  

\subsection{Collider searches for particles with LNV interactions} 

The specific models above also have different collider signatures:
the LR model contains heavy fermions and scalars, while the RR
model contains only heavy scalars; all of which can be searched for at
the LHC. Verification of either model would involve not only
the discovery of the corresponding heavy particles, but 
also a demonstration of the presence of 
LNV interactions.
(the possibility of observing 
LNV events at LHC was emphasized quite some time ago \cite{Keung:1983uu}). 
Although not all decays of the new particles produce LNV signals, 
in some cases they may be dominant; however, in general one has to search 
for the new resonances in the most sensitive channels, and only afterwards
address the possibility of observing LNV events. 
Typically, these will be difficult to observe because LNV is usually a small
effect since
the corresponding amplitudes involve
several (small) couplings all of which must be present in order
for LN to be broken. 
Hence, in general the dominant production mechanisms are standard and 
LN conserving, otherwise they are small
(though LNV decays can be slow, they typically still can occur
within the detector \cite{Franceschini:2008pz}).

Following the discussion in Ref. \cite{delAguila:2011gr} we will 
first comment on the detection of the new scalars in the RR model. 
Doubly-charged scalars have fixed couplings to photons, so
that their production cross section is known;
their decay into leptons (if allowed) gives a very clean signal, 
which is particularly important at hadronic machines; 
although it is not LNV by itself. 
Therefore, if doubly-charged scalars are light enough, they are
quite suited for detection at colliders. 
Generally, this type of scalars forms part of a weak triplet, 
and usually also acts as see-saw messenger of type II 
 \cite{Konetschny:1977bn,Cheng:1980qt,Lazarides:1980nt,Magg:1980ut,Schechter:1980gr,Mohapatra:1980yp,Gelmini:1980re} (see also
\cite{Ma:1998dx}).
These triplets are then theoretically well-motivated, especially
when considering L-R symmetric models, and
simulations of their production at future colliders 
can be found in the literature 
\cite{Gunion:1989in,Huitu:1996su,Gunion:1996pq,Akeroyd:2005gt,Azuelos:2005uc}
(see also \cite{delAguila:2008cj,Akeroyd:2010ip} for recent studies;
and for model-independent ones \cite{Dion:1998pw,Cuypers:1996ia}). 
The general conclusion is that the LHC discovery limit can reach masses
over $600$ GeV (for a CM energy of $14$ TeV and an integrated 
luminosity of 30 fb$^{-1}$) \cite{delAguila:2008cj,delAguila:2009bb}; 
although the actual limits may be larger given 
the outstanding LHC performance for a CM energy of $7$
TeV~\cite{delAguila:2010uw}.  
(See for a recent review \cite{Nath:2010zj}.) 
First results from CMS \cite{cms-pas-hig-11-007:2011} and ATLAS~\cite{atlas-conf-2011-127}
have been presented at this last CM energy with an integrated 
luminosity of $0.89\,\mathrm{fb^{-1}}$ for any di-lepton scalar final state 
and of 
 $1.6\,\mathrm{fb^{-1}}$ for di-muon final states, respectively. 
As nothing is seen, a lower bound on the doubly-charged scalar mass is obtained:
of about $250\,\mathrm{GeV}$ if the main decay
channels contain $\tau$ leptons and to about $300\,\mathrm{GeV}$ 
if they contain only electrons or muons \cite{cms-pas-hig-11-007:2011}, 
reaching $375\,\mathrm{GeV}$  if they only couple 
to muons~\cite{atlas-conf-2011-127}. 
(Present Fermilab Tevatron Collider limits are less
stringent~\cite{Abazov:2011xx,CDF-PHYS-EXO-PUBLIC-10509}.) 
In our case, however, the 
doubly-charged triplet, $ \chi^{\pm\pm}$ does not directly couple to fermions;
while the other 
doubly-charged scalar in the model, a singlet $ \kappa^{\pm\pm}$, does not
couple to $W$ pairs. 
However, they mix; and 
both of them can be produced at LHC via the Drell-Yan mechanism 
($q\bar{q}\rightarrow\gamma^*, Z^*\rightarrow \chi^{++}\chi^{--},
\kappa^{++}\kappa^{--}$). 
Since this  is the main production process assumed
by both LHC Collaborations, the former limits apply directly  to the singlet
decaying dominantly to lepton 
pairs for a small mixing:
$m_{\kappa} > 300\,{\mathrm {GeV}} $. Limits on the triplet mass 
will be more difficult to derive, for the dominant process  
$q\bar{q}\rightarrow \gamma^*, Z^*\rightarrow \chi^{++}\chi^{--}\rightarrow
W^+W^+W^-W^-$ 
is more complicated to study, due to its large backgrounds and the 
inherent difficulty of reconstructing several leptonic $W$ decays \cite{Chiang:2012dk}. 
The only viable LNV decay channels 
$q\bar{q}\rightarrow\gamma^*, Z^*\rightarrow \chi^{++}\chi^{--},
\kappa^{++}\kappa^{--} 
\rightarrow e_a^\pm e_b^\pm W^\mp W^\mp$ are suppressed by 
the small mixing between $\chi^{\pm\pm}$ and $\kappa^{\pm\pm}$. 

The LR model discussed above contains, besides the scalar isotriplet of unit 
hypercharge $\chi$, a second scalar isodoublet and several vector-like lepton
doublets. 
This scalar triplet couples to a light and a heavy lepton, and the latter 
decays into a light lepton and a $W$, or a $Z$, or a Higgs boson; this results
in a four-fermion decay. Similarly, the extra scalar isodoublet 
decays into four fermions. 
Then, if also pair produced via the Drell-Yan mechanism, final states will have
at least 
eight fermions. The signal may be striking due to the large number of charged
leptons, 
but there are many open channels and may be not easy to resolve the different
samples. 
These final states are different from those of the RR model, which will  
eventually allow to discriminate between both theories. 

The heavy vector-like lepton doublets are also mainly produced in pairs. They
violate 
the Glashow-Iliopoulos-Maiani mechanism \cite{Glashow:1970gm}, and can decay 
through a flavor changing neutral current into a light lepton and a $Z$ or Higgs
boson 
\cite{delAguila:1989rq}; as for sequential fermions
they can also decay into a lepton and a $W$ boson
through the usual charged current interaction.
In these cases the final states have at least six fermions, so that 
the heavy leptons will be
relatively easy to find if light enough: $M_a \lesim$ TeV 
for a CM energy of $14$ TeV and an integrated luminosity of 100 fb$^{-1}$
\cite{delAguila:1989rq,AguilarSaavedra:2009ik,delAguila:2010es}. 
The dominant decays of these (quasi) Dirac fermions 
are LN conserving to a large extent.

\section{Conclusions}

To date there is no direct evidence that LN is not a
symmetry of nature \cite{Nakamura:2010zzi}, at least at the energies 
and experimental sensitivities currently available. 
There are, however, scenarios that suggest LN may be violated
at higher scale. For example, the baryon asymmetry
of the universe (see for example \cite{Dolgov:1993qm}) can be explained through
leptogenesis \cite{Fukugita:1986hr}, yet  no   
experimental indication of such a mechanism has been observed.
Light neutrino masses and mixing angles provide the
most favored explanation for neutrino oscillations, but
the experiments do not distinguish between Dirac and 
Majorana masses, and thus, they do not require LNV;
although, in many models light neutrinos are assumed to be
Majorana fermions, implying that LNV is also assumed.

In contrast \znbb decay 
and appropriate signals at LHC will be sensitive to LNV effects; and in fact, these
are the only way known to 
experimentally establish the presence of LNV. 
Hence the relevance of these 
two types of experiments. 
A new generation of \znbb decay experiments are 
underway (see \cite{Barabash:2011fg,Avignone:2007fu} for recent reviews), where 
a positive signal would provide conclusive 
evidence of LNV, and would open a new era of 
experimental searches and theoretical studies aiming at
isolating the type of NP that mediates such a process.

The effective Lagrangian approach allows to address this theoretical 
question with generality and has  been the subject of the above discussion. 
As described in the text, and detailed in the appendix, 
we have constructed all gauge-invariant 
effective operators of dimension $\le9$,  violating LN by 2 units
and involving two SM leptons (but no quarks) 
and any number of Higgs isodoublets and covariant 
derivatives (the appendix also lists LNV operators 
with more than 2 leptons). For each of the three possible scenarios;
{\it(i)} with two LH leptons (LL), {\it(ii)} one LH 
lepton and one RH electron (LR), and {\it(iii)} two RH electrons (RR), there 
is only one lowest order effective operator, $\ocal\up5, \ocal\up7$ 
and $\ocal\up9$, respectively 
(see Eqs. (\ref{eq:lowestoperatorLL}), (\ref{eq:lowestoperatorLR}) and 
(\ref{eq:lowestoperatorRR})). 
They describe the largest possible contribution to \znbb decay 
for each final electron chirality assignment. 
We have also identified the possible new particle additions 
that can generate these operators at tree level
(assuming that the full theory is renormalizable).

In general, models of neutrino masses can have an origin different from that
of \znbb decay; irrespective of that,
once the NP generates any of the LNV operators, the theory will generate 
neutrino masses at some loop order: at tree level, at one loop or at two 
loops, depending on whether the operator is $\ocal\up5, \ocal\up{\oiv}$ or 
$\ocal\up9$, respectively. 
For $ \ocal\up{7,9}$ the corresponding masses are
relatively 
suppressed by loop and light mass (generated by chirality flips) factors. 
In the case of $\ocal\up5$ there are no such suppressions and
the same parameter giving the effective neutrino 
electron mass $({m_\nu})_{ee}$ enters in the \znbb decay 
amplitude (see Eq. \eqref{eq:amplitude0nu2betaLL}).
Thus, both pieces of data only constrain the $|C\up5|/\Lambda$ ratio.
In contrast in the other two cases, and since the neutrino mass scale is
fixed to be around $0.1$ eV, the observation of \znbb
decay would allow to estimate the scale $\Lambda$ of NP and the
corresponding effective operator coefficient $C$ for each scenario
for natural theories (and perturbative couplings).
Thus, leaving to experiments searching for LFV and for collider
signatures of LNV mediators.
To provide an existence proof we have also constructed a simple,
realistic model for each scenario, explicitly calculating the neutrino
masses;
as expected from dimensional arguments these masses are always proportional
to $1/\Lambda$ (see Eqs. \eqref{eq:o5.mnu}, \eqref{eq:o7mnu}, \eqref{eq:massesRR},
\eqref{modelneutrinomass}  and Ref.~\cite{delAguila:2011gr}),
as argued in the introduction: Eqs. \eqref{al:massesLLcorrect}, \eqref{al:massesLRcorrect} and 
\eqref{al:massesRRcorrect}. 
Then, whereas $C\up5$ and its associated $\Lambda$ can vary in between
eleven orders of magnitude, at least $C\up7_{\tau\tau}$ ($C\up9_{\tau\tau}$) must be at the per million
(mille) level and the corresponding $\Lambda$ of the $10$~(few)~TeV order.
Besides, any spectrum of neutrino masses within present experimental
limits can be accommodated in the first two cases, but in the RR model the
neutrino hierarchy must be normal and the third mixing angle
$\sin^2 \theta_{13} \gtrsim 0.008$,
in agreement with recent observations
\cite{Adamson:2011ig,Abe:2011sj,Abe:2011fz,An:2012eh,Ahn:2012nd}.
On the other hand, once an explicit model is at hand, we can 
check if it does satisfy the present LFV constraints and 
bounds from large colliders; this 
is the case for the models discussed. Although 
the Higgs searches now underway at LHC will stringently 
restrict these models with an extended scalar sector 
near the electroweak scale.

\section*{Acknowledgements}

This work has been supported in part by the Ministry of Economy and 
Competitiveness, under the grant numbers FPA2006-05294,  FPA2010-17915
and FPA2011-23897, by the Junta de Andaluc{\'\i}a 
grants FQM 101, FQM 03048 and FQM 6552, by the 
``Generalitat Valenciana'' grant PROMETEO/2009/128, and by the 
U.S. Department of Energy grant No.~DE-FG03-94ER40837. 
A.A. is supported by the MICINN under the FPU program. 

\section*{Appendix: Effective Lagrangian description of NP}
\label{sec:app1}

In this appendix we provide for completeness a brief summary of
the effective Lagrangian approach and describe the procedures we
followed in constructing the operators discussed in the
main text.

Effective theories are useful for situations where there is a
scale gap: some type of heavy physics effects contribute only
virtually since the available energies are well below the scale
of these interactions. Technically one differentiates between
the case where the underlying physics decouples \cite{Appelquist:1974tg} and when it
does not. For the first case the low-energy effective theory is
obtained by a formal expansion in inverse powers of the heavy
scale, the existence of which is guaranteed by the decoupling
theorem \cite{Appelquist:1974tg} (see also \cite{Collins:1984xc}). 
For non-decoupling heavy physics the effective
theory is obtained as a derivative expansion \cite{Weinberg:1978kz}. 

Here we shall 
assume that the heavy physics is both decoupling and weakly
coupled, so that a perturbative expansion is appropriate;
the characteristic scale of these new interactions will
be denoted by $ \Lambda $.
The general parameterization of NP effects using
effective interactions is valid at energies below
$ \Lambda $. The procedure is straightforward: one 
constructs all Lorentz-invariant operators involving
the light fields and their derivatives and respecting
the low-energy local symmetries (here, the SM
gauge symmetries); the effective Lagrangian is
then the linear combination of all
such operators, where the $\Lambda$-dependent
coefficients parametrize all possible
(weakly-coupled and decoupling) types of heavy 
physics. In our case the NP will violate LN 
but will not couple to quarks. 
If the theory underlying the SM were known then
one could derive the low-energy effective theory 
coefficients in terms of the parameters of the model.
It may then happen that some operators will be absent or may appear
with suppressed coefficients due to some as yet unknown
symmetry. Not knowing the correct SM extension,
the effective Lagrangian coefficients are treated as unknowns
susceptible to experimental determination.

The coefficient of an operator of (canonical) dimension $n$ 
is proportional to $ \Lambda^{4-n} $ so that the 
larger the dimension of the
operator the smaller its effect; given a finite experimental
precision this implies that operators with $n$ sufficiently
large can be ignored. In addition, operators that are generated 
by heavy particle loops have coefficients that receive
a typical loop suppression factor $ \sim 1/(4\pi)^2 $.
It is important to note that any operator
that respects the local symmetries of the SM will be
generated by the NP at some loop level. Whether this happens
at tree level depends on the operator and the details of the theory
underlying the SM; it is a simple exercise to determine the 
operators and types
of NP that have this property. It is also easy to device 
types of NP for which all tree-level-generated operators (TLGOs) are 
absent~\footnote{For example, if there is a discrete symmetry under 
which all the SM particles are singlets but none of the new particles are
\cite{Wudka:2005yp}.}.
Dominating effects are then associated with the lowest-dimension
TLGOs contributing to the process at hand. 
In this it
is important to note that whether an operator is generated at tree level
or not depends on the details of the heavy physics; there are,
however, operators that are necessarily loop generated
by {\em all} modalities of heavy physics \cite{Arzt:1994gp}; 
we call these loop-generated operators (LGOs). It is also
worth keeping in mind that in most cases the effects
from LGOs will compete with 
those generated radiatively by the SM and are often
subdominant; exceptions occur when the SM effects are
absent due to some accidental SM symmetry, such as 
custodial symmetry or LN. Except for
these cases LGO effects lie beyond
the experimental sensitivity of current experiments~\footnote{
This means that radiative effects generated by the heavy
excitations are too small to be observed, it does {\em not}
preclude direct observation of new particles (provided
the energy available is high enough). In these cases
the effective Lagrangian approach is, however, inapplicable.}. 
There is one additional observation that can be used
to simplify the effective Lagrangian: if two operators
$ \ocal $ and $ \ocal'$ are such that 
the combination
$\ocal - \ocal' $
vanishes on-shell (that is, when the classical equations of motion
are imposed), then the $S$-matrix 
depends only on
the sum of the corresponding operator coefficients
 \cite{Georgi:1991ch,Arzt:1993gz},
so that one of the operators can be omitted from the
effective Lagrangian parameterization~\footnote{Note in 
particular that this result implies that the experimental
sensitivity to $\ocal$ and $\ocal'$ is the same: one cannot replace
an operator $ \ocal' $ by an equivalent one $ \ocal $ and
find weaker limits on $\ocal$ than on $\ocal' $.}.

As emphasized previously, the main feature of the processes we
will be interested in is that they exhibit LNV. The 
operators of interest have dimensions $\geqslant 5$; 
many of them have been enumerated in earlier
publications, including operators involving
quarks~\cite{Babu:2001ex,Choi:2002bb,Engel:2003yr,deGouvea:2007xp}, 
though such catalogs are not exhaustive.
We do not pretend to provide a complete list of operators
but concentrate instead on those low-dimensional TLGOs 
with two leptons and any number of bosons and derivatives 
that can be probed experimentally and provide leading
effects for wide classes of interesting heavy physics.
In the following we will assume that the
low-energy excitations are those of the SM, 
with a scalar sector containing $n_d \ge1$
doublets $\phi_i, ~i=1, \ldots, n_d$
of hypercharge $1/2$. 
Although generally we assumed 
only one (Higgs) isodoublet when classifying the 
possible higher-order operators,  
it is worth considering the case $n_d > 1 $
because this allows the presence of additional
operators that are absent in the one-doublet
case for symmetry reasons; we provide examples
of this in the tables below.
Moreover, as we illustrated in Section \ref{sec:LR-model},
the introduction of 
extra scalar doublets
may allow for simpler, phenomenologically viable, fundamental 
theories.
In contrast, no additional light 
fermions are assumed to exist, in particular, any 
RH gauge-singlet fermions (such as RH 
neutrinos) are assumed to be heavy \footnote{
Though it will not be considered in this
appendix, it is
straightforward to extend the light scalar
sector by adding a number of light scalar gauge-singlets.}. 
In presenting our expressions we will not
display family indices, though in general these are present. 

In order to simplify the notation through this appendix the LH lepton isodoublets 
are denoted by $ \ell $ and
RH lepton isosinglets by $e$; here we will consider
operators involving only leptons, those involving quarks will
be discussed in a future publication. 
We now provide the list of LNV operators of dimension
$ \le 9 $. In doing so we merely
provide field content, with the understanding that all possible
gauge and Lorentz contractions are to be counted
(so that each of the entries represents, in general, 
more than one operator).  
Also to make the notation simpler and clarify the 
physical effects of the operators we find it convenient to 
introduce the following composite operators 
\beq
 N_{ia} = \phid_i \lt_a \,, \qquad 
 \Psi^\mu_{ia}= \phid_i D^\mu \lt_a \,, \qquad 
J^\mu_{ab} = \lb_a D^\mu \lt_b \,, \qquad
\tilde\wcal^\mu_{ij} = \phid_i D^\mu \phit_j \,,
\eeq
where, as before,
\beq
\ell = {\bpm \nu_L \cr e_L  \epm} \,, \qquad 
\lt = \epsilon C \overline{\ell}^T \,, \qquad
\phit_i = \epsilon \phi_i^*
\eeq
with $ \epsilon = i \sigma_2$. Lower-case indices $a,b$, etc. 
are family indices. 
For the case of a single leptonic  family and one scalar isodoublet with
$ \vevof\phi = ( 0 , v)^T $ the above operators become
\bea
N &=& - v \nu_L^c  + \cdots \,, \cr
\Psi^\mu &=& - v \left[ \left( \partial_\mu + \frac i2 \frac g{\cw} Z_\mu
\right) \nu_L^c
+ \frac {i g}{\sqrt{2}} W_\mu^- e_L^c \right] + \cdots \,, \cr
J_\mu &=& \overline{\nu_L} \left[ {\stackrel \leftrightarrow{\partial_\mu}} 
 - i g (\cw Z_\mu + \sw A_\mu) \right] e_L^c
+ \frac{ig}{\sqrt{2}} \left( W^+_\mu \overline{\nu_L} \nu_L^c - 
W_\mu^- \overline{e_L} e_L^c
\right) 
+ \cdots \,, \cr
\wcal^\mu &=& -i \frac g{\sqrt{2}} v^2 W^-{}^\mu + \cdots \,,
\label{eq:expan}
\eea
where the ellipsis denote terms involving the physical scalars
and we defined $\nu_L^c = C \overline{\nu_L}^T $; numerically
$ v \simeq 174~\gev $.

Given an operator $ \ocal $ it is straightforward to determine whether
it is loop generated or whether there are models where it appears
at tree level \cite{Arzt:1994gp}. For the second case one can also determine the
types of heavy excitations involved in generating $ \ocal $, of which
there will be in general several possibilities. Below
we will also provide the tree-level diagrams and the
list of heavy excitations that can generate the leading LNV operators
involving two leptons (for a systematic study of the dimension 5 operator at one loop see~\cite{Bonnet:2012kz}). As in the text, in 
listing the heavy particles we denote by
$\vec IY,~ \fer IY $ and $ \sc IY $
a heavy vector, fermion or scalar with isospin
$I$ and hypercharge $Y$, respectively. When the
heavy particles can be either a heavy
vector or heavy scalar with the same
isospin and hypercharge, we use $ \bos IY$
to denote both possibilities.

\subsection{LNV operators with no quarks}
\label{sec:l-e.LNV}

All the operators below can be generated at tree level,
and it is a simple exercise to determine the types of heavy
physics that can do so.
The number of possibilities, however, increases rapidly
with the dimension of the operator so we will
restrict ourselves to those with the lowest dimension
operators within each group. Still it is
useful to note the following: if $ \ocal$ differs
from $ \ocal' $ by the presence of 2 derivatives,
$ \ocal \sim D^2 \ocal' $, then if $\ocal$ is
generated at tree level in a certain model,
$ \ocal'$ will be also generated at tree level.
Thus, for example, if a model generates a low-energy
effective vertex $ \sim e_L^2 W^2 $ (from
$D^2 \ell^2 \phi^2$), it will also generate
a Majorana-mass term $ \nu_L^2 $ at tree level.
In general, however, it will {\em not} generate a
$e_R^2 W^2 $ vertex at tree level. 
As mentioned repeatedly, this particular
example is of interest when studying
\znbb decay.

Below we list the LNV operators of dimension $\le 9$ not involving
quarks together with the sets of heavy excitations that may generate
them at tree level. 
The operators involving two leptons can be grouped in 3 sets according
to the chirality of the light leptons
(operators with $>2$ leptons will be listed at the end,
and are provided for completeness only):
\paragraph{Two LH leptons plus bosons}
\beq
\begin{array}{|c|l|l|}
\hline
\hbox{Dim.} & \multicolumn{1}{|c|}{\hbox{Operator(s)}} \cr \hline
5 & \ell^2 \phi^2   \cr
7 & \ell^2 \phi^2 \Omega   \cr
9 & \ell^2 \phi^2 \Omega^2  
\cr\hline
\end{array}
\eeq
Where $\Omega$ denotes either $D^2$ or $ \phi \phit $:
\beq
\Omega \sim (D^2, \phi \phit ) \,. 
\eeq
So that, for example, $\ell^2 \phi^2 \Omega $
corresponds to $\ell^2 \phi^3 \phit $
or $D^2 \ell^2 \phi^2$. As mentioned above, each
entry represents a series of operators obtained
by making all possible index contractions and
having the derivatives operate on all fields.
Considering all possible contractions the 
operators of dimension 5 are simply
\beq
\ocal^{(5)\dagger}_{ijab} = N^T_{ia} C N_{jb} \,. 
\label{eq:ll5}
\eeq
The operators of dimension 7 in this category are
\bea
\ocal\up{\oi}_{ijab} &=& \Psi^T_{ia} \cdot C \Psi_{jb} \,, \cr
\ocal\up{\oii}_{ijab} &=& J_{ab} \cdot \tilde\wcal_{ij} \,, \cr
\ocal\up{\oiii}_{ijab} &=& N_{ia}^T C \partial_\mu \Psi^\mu_{jb} \,, 
\label{eq:ll7}
\eea
as well as $ \ocal\up5_{ijab} (\phi_k^\dagger \phi_l) $. 
We will not consider operators of dimension 9 in this category. 

The heavy excitations that can generate (\ref{eq:ll5})
and (\ref{eq:ll7}) at tree level can be read off Table
\ref{tab:a1} and are listed in (\ref{O5mediators}).
\begin{table}
$$
\begin{array}{|c|c|c|c|c|}
\hline
\multicolumn{5}{|c|}{\includegraphics[width=1.5in]{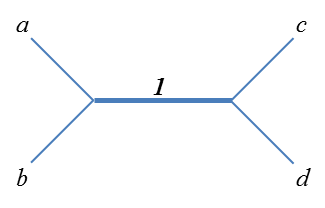}} \cr\hline
1                  & a    & b    & c    & d    \cr \hline\hline
\sc11 \td          & \ell & \ell & \phi & \phi \cr\hline
\fer00 \aut \fer01 & \ell & \phi & \ell & \phi \cr\hline
\end{array}
$$
\caption{Diagrams generating $\ocal\up5$ at tree level;
fields in smaller font inside brackets correspond to 
models with more than one light scalar isodoublet.}
\label{tab:a1}
\end{table}
One may think that (\ref{eq:ll7}) are still relevant in that they involve
also the gauge bosons, but in actual calculations 
{\em any} amplitude involving these
effective operators will have a counterpart involving (\ref{eq:ll5})
together with SM vertices. The second amplitude will be  suppressed
by only $1/ \Lambda $, compared to $1/\Lambda^3$ for the first one,
with the {\em same} scale in both cases. As a result
{\em all} effects of (\ref{eq:ll7}) are subdominant in any process;
accordingly, 
we ignore these operators in the following. The same holds
for operators of the form $\ocal\up5 (\phi^\dagger\phi)$,
and for operators of dimension 9 in this category.

\paragraph{One RH and one LH lepton plus bosons}
\beq
\begin{array}{|c|l|}
\hline
\hbox{Dim.} & \multicolumn{1}{|c|}{\hbox{Operator(s)}} \cr \hline
7 & D e \ell \phi^3   \cr
9 & D e \ell \phi^3 \Omega  \cr \hline
\end{array}
\eeq
The leading operator of this class  has dimension 7 and is given by 
\beq
\ocal\up{\oiv}_{ijkab} =  \tilde\wcal_{ij}^\mu (\overline{e_a} \gamma^\mu N_{kb})
=  i  x_i x_j x_k \mw v^2 \overline{e_a} \not\!\! W^- P_R \nu_b^c + \cdots \,, 
\eeq
where the ellipsis denote terms involving the physical scalars and
\beq
\vevof{\phi_i} = v x_i \,, \quad \sum_i x_i^2 = 1 \, , \qquad
m_W^2 = \half g^2 v^2
\eeq
with $g$ the $\su2$ gauge-coupling constant.

In order to determine the types of heavy excitations
that generate $ \ocal\up\oiv$ we proceed as follows.
We first write (omitting subindices for simplicity)
\beq
\ocal\up{7} = \left(\bar e \gamma^\mu \phi^\dagger
\tilde \ell\right)
\left(\phi^\dagger \partial_\mu \phit \right) + \cdots \,, 
\label{eq:o7-1}
\eeq
where the ellipsis denote terms with vector bosons.
This term will be generated by graphs containing two
charged leptons of opposite chiralities and 3 light
scalars as external legs. A model that
generates this graph at tree level must also
generate the full operator (at tree level) due to gauge invariance.
In practice this is obtained by attaching an appropriate
number of light gauge boson
lines to the internal heavy propagators at all points
allowed by the quantum numbers.
The graphs that can generate the term (\ref{eq:o7-1}) at tree
level can have two topologies, but chirality prevents one of them;
the remaining tree-level graphs are given in Table \ref{tab:a2}
and listed in (\ref{eq:IV7H}).
\begin{table}
$$
\begin{array}{|c|c|c|c|c|c|c|}
\hline
\multicolumn{7}{|c|}{\includegraphics[width=1.5in]{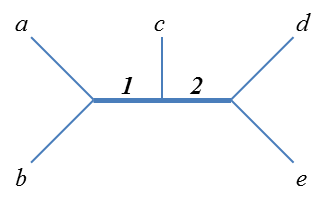}} 
\cr\hline
1              & 2     & a & b & c    & d    & e  \cr\hline\hline
\vec{1/2}{3/2} & \vec01 \aut \vec11 & \ell & e    & \phi & \phi & \phi \cr\hline
\vec{1/2}{3/2} & \sc11  \td         & \ell & e    & \phi & \phi & \phi \cr\hline
\fer10         & \vec11             & \ell & \phi & e    & \phi & \phi \cr\hline
\fer10  \tdf   & \sc11  \td         & \ell & \phi & e    & \phi & \phi \cr\hline
\fer00         & \vec01             & \ell & \phi & e    & \phi & \phi \cr\hline
\fer{1/2}{1/2} & \vec01 \aut \vec11 & e    & \phi & \ell & \phi & \phi \cr\hline 
\fer{1/2}{1/2} & \sc11  \td         & e    & \phi & \ell & \phi & \phi \cr\hline 
\fer{1/2}{1/2} & \fer00 \aut \fer10 & e    & \phi & \phi & \phi & \ell \cr\hline
\end{array}
$$
\caption{Diagrams generating $\ocal\up\oiv$ at tree level;
fields in smaller font inside brackets correspond to 
models with more than one light scalar isodoublet.}
\label{tab:a2}
\end{table}

\paragraph{Two RH leptons plus bosons}
\beq
\begin{array}{|c|l|}
\hline
\hbox{Dim.} & \multicolumn{1}{|c|}{\hbox{Operator(s)}} \cr \hline
7 & e^2 \phi^4   \cr
9 & e^2 \phi^4  \Omega  \cr \hline
\end{array}
\eeq
The operators $ e^2 \phi^4 $ and $e^2 \phi^5 \phit$ vanish
for the case of a single scalar doublet. When there are more
scalar isodoublets these operators generate 
vertices of the form $ e e H^+ H^+ $ multiplied by
neutral scalars and/or \vev s. 
Barring the presence of light single-charged scalars the
leading operator of this class then has dimension 9 and is given by
\beq
\ocal\up9_{ijklab} = \bar e_a e^c_b \tilde\wcal_{ij} \cdot \tilde\wcal_{kl}\, .
\eeq
The tree-level graphs that can generate this operator are obtained
in the same way as above. We expand
\beq
\ocal\up9 = (\bar e e^c) \left(\phi^\dagger \partial^\mu \phit \right)
\left(\phi^\dagger \partial_\mu \phit \right) + \cdots \,, 
\eeq
where the ellipsis denote terms with vector bosons and then look
for graphs with two RH electrons and 4 scalars in the external lines, which
generate
this  term in $ \ocal\up9$; the rest of the operator will necessarily 
be generated because we assume the underlying theory respects the SM 
gauge symmetry.
There are three diagram topologies presented in Tables \ref{tab:a3I},
\ref{tab:a3II} and \ref{tab:a3III}.
The sets of heavy excitations that can generate the dimension 9 operators 
at tree level are listed in (\ref{eq:O9.NP}). 
\begin{table}
$$
\begin{array}{|c|c|c||c|c|c|c|c|c|}
\hline
\multicolumn{9}{|c|}{\includegraphics[width=1.5in]{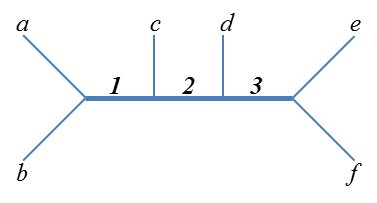}} \cr\hline
1              & 2                  & 3                  & a & b & c    & d    &
e    & f    \cr\hline\hline
\sc02          & \bos{1/2}{3/2}     & \sc11 \td          & e & e & \phi & \phi &
\phi & \phi \cr\hline
\sc02          & \bos{1/2}{3/2}     & \vec11 \aut \vec01 & e & e & \phi & \phi &
\phi & \phi \cr\hline
\fer{1/2}{1/2} & \bos{1/2}{3/2}     & \sc11  \td         & e & \phi & e & \phi &
\phi & \phi \cr\hline
\fer{1/2}{1/2} & \bos{1/2}{3/2}     & \vec11 \aut \vec01 & e & \phi & e & \phi &
\phi & \phi \cr\hline
\fer{1/2}{1/2} & \fer10 \tdf        & \sc11  \td         & e & \phi & \phi & e &
\phi & \phi \cr\hline
\fer{1/2}{1/2} & \fer10             & \vec11             & e & \phi & \phi & e &
\phi & \phi \cr\hline
\fer{1/2}{1/2} & \fer00             & \vec01             & e & \phi & \phi & e &
\phi & \phi \cr\hline
\fer{1/2}{1/2} & \fer00 \aut \fer10 & \fer{1/2}{1/2}     & e & \phi & \phi & e &
\phi & \phi \cr\hline
\bos11         & \fer10 \tdf        & \sc11 \td          & \phi & \phi & e & e &
\phi & \phi \cr\hline
\vec11         & \fer10             & \vec11             & \phi & \phi & e & e &
\phi & \phi \cr\hline
\vec01         & \fer00             & \vec01             & \phi & \phi & e & e &
\phi & \phi \cr\hline
\end{array}
$$
\caption{Diagrams generating $\ocal\up9$ at tree level with three virtual
particles;
fields in smaller font inside brackets correspond to 
models with more than one light scalar isodoublet.}
\label{tab:a3I}
\end{table}
\begin{table}
$$
\begin{array}{|c|c|c||c|c|c|c|c|c|}
\hline
\multicolumn{9}{|c|}{\includegraphics[width=1.5in]{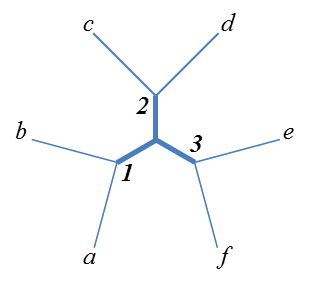}} \cr\hline
1              & 2                    & 3              & a & b & c    & d    & e
   & f    \cr\hline\hline
\sc02          & \sc11  \td           & \sc11  \td     & e & e & \phi & \phi &
\phi & \phi \cr\hline
\sc02          & \sc11  \td           & \vec11 \tdv    & e & e & \phi & \phi &
\phi & \phi \cr\hline
\sc02          & \vec11               & \vec11         & e & e & \phi & \phi &
\phi & \phi \cr\hline
\sc02          & \vec01               & \vec01         & e & e & \phi & \phi &
\phi & \phi \cr\hline
\fer{1/2}{1/2} & \sc11  \td           & \fer{1/2}{1/2} & e & \phi & \phi & \phi
& \phi & e \cr\hline
\fer{1/2}{1/2} & \vec11 \aut \vec01   & \fer{1/2}{1/2} & e & \phi & \phi & \phi
& \phi & e \cr\hline
\end{array}
$$
\caption{As table~\ref{tab:a3I} but with a different topology.}
\label{tab:a3II}
\end{table}
\begin{table}
$$
\begin{array}{|c|c||c|c|c|c|c|c|}
\hline
\multicolumn{8}{|c|}{\includegraphics[width=1.5in]{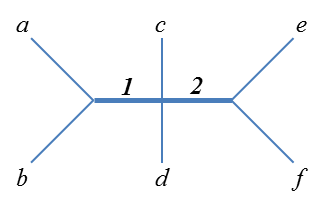}} \cr\hline
1              & 2         & a & b & c    & d    & e    & f    \cr\hline\hline
\sc02          & \sc11 \td & e & e & \phi & \phi & \phi & \phi \cr\hline
\end{array}
$$
\caption{As table~\ref{tab:a3I} but with only two virtual particles.}
\label{tab:a3III}
\end{table}

\paragraph{More than 2 leptons and bosons}

\beq
\begin{array}{|c|lll|}
\hline
\hbox{Dim.} & \multicolumn{3}{|c|}{\hbox{Operator(s)}} \cr \hline
7 & \ell^3 e^c \phi   & &\cr
9 & \ell^3 e^c \phi \Omega   & \quad
( D \ell^2 \phi^2 , \;  \ell e \phi^3) \times(\ell \lt, e e^c) & \quad \ell^4
(e^c)^2 \cr
\hline
\end{array}
\eeq
where the middle term in the last line represents objects of the form
$ D \ell^3 \lt \phi^2 ,\, \ell^2 \lt e \phi^3$, etc.

\bigskip

The important conclusion of the above arguments is that there are only
three interesting LNV operators of lowest dimension: $\ocal\up5,~
\ocal\up\oiv,~\ocal\up9$. Aside form their LNV effects
these operators in general transform non-trivially
under CP.

%\bibliographystyle{myuphys}
%\bibliographystyle{JHEP}
%\bibliographystyle{myutphysnew}

%\bibliography{tot-ins-bis}

\providecommand{\href}[2]{#2}\begingroup\raggedright\endgroup

\end{document}